\documentclass[aps,prb,amsmath,twocolumn,amssymb,titlepage,superscriptaddress]{revtex4-2}
\usepackage[english]{babel}
\usepackage{graphicx} 
\usepackage{transparent}
\usepackage{amsmath}
\usepackage{amssymb}
\usepackage{epstopdf}
\usepackage{amsfonts}
\usepackage{bm}
\usepackage{times}
\usepackage{color}
\usepackage{mathtools}
\usepackage{hyperref}
\usepackage{tabularx}
\usepackage{hhline}

\newcommand{\ket}[1]{|#1\rangle}
\newcommand{\bra}[1]{\langle#1|}

\newcommand{\astcycl}{\mathrlap{\kern0.085em{\circlearrowright}}\ast}
\newcommand{\taucycl}{\mathrlap{\kern0.42em{\bullet}}\circlearrowright}

\begin{document}

\title{High order strong-coupling expansion for X-ray absorption on a dynamically screened impurity}
\author{Eva Paprotzki}
\affiliation{I. Institute for Theoretical Physics, University of Hamburg, Notkestraße 9-11, 22607 Hamburg, Germany}
\affiliation{The Hamburg Centre for Ultrafast Imaging, Hamburg, Germany}
\author{Martin Eckstein}
\affiliation{I. Institute for Theoretical Physics, University of Hamburg, Notkestraße 9-11, 22607 Hamburg, Germany}
\affiliation{The Hamburg Centre for Ultrafast Imaging, Hamburg, Germany}

\begin{abstract}
Time-resolved X-ray absorption can reveal the dynamical screening of the local Coulomb interaction in strongly correlated  photo-excited materials. Here, we focus on the theoretical prediction of X-ray absorption in the presence of dynamical screening using the strong coupling expansion, i.e., an expansion around the isolated absorption site in terms of the retarded interaction. The evaluation of higher order diagrams is made numerically feasible by an approach based on the decomposition of the retarded interaction into complex exponentials. With this, we evaluate the strong coupling series to third order on an electron-boson model of Holstein type. We demonstrate that in relevant coupling regimes, even low orders of the strong coupling expansion can give a significant correction over the previously used lowest order approximation.
\end{abstract}

\maketitle

\section{Introduction}
\label{sec:intro}

Time-resolved X-ray spectroscopies  can now reach sub-eV resolution \cite{Cao2019, Gerasimova2022}. This offers promising pathways to investigate solids out of equilibrium \cite{Sentef2021,Giannetti2016,Murakami2023}, utilizing both time-resolved X-ray scattering and time-resolved X-ray absorption spectroscopy (XAS)  \cite{Mitrano2020,Mitrano2024}. A notable example includes the shifts in the XAS absorption lines of laser-excited transition metal compounds  \cite{Baykusheva2022,Lojewski2024,graanas2022,wang2022ultrafast}, which can signify a transient renormalization of the local Coulomb interaction \cite{Golez2019,Tancogne2018} due to dynamical screening. 

In general, interpreting time-resolved XAS signals requires a microscopic understanding \cite{Wang2020, Chen2019, Werner2022}. The final state of X-ray absorption in correlated materials often involves a bound exciton formed from the valence excitation and the core hole. As a result, XAS can often be understood using cluster models \cite{Gunnarsson1983} that include a single absorption site. Similarly, in dynamical mean-field theory (DMFT), the XAS signal is computed from the DMFT impurity model supplemented with a core orbital \cite{Cornaglia2007,haverkort2014, luder2017, hariki2018,Werner2022}. Within such an impurity-based description, dynamical screening is captured by coupling the charge on the impurity to dynamic charge fluctuations that are represented by a continuum of bosonic modes. These screening modes mediate a retarded local interaction, which can lead to both line shifts and satellites of the XAS lines. For instance, in DMFT+GW calculations \cite{Biermann2003,Sun2002}, which represent the screening by an additional bosonic environment within the impurity model, XAS line shapes are modified after photo-excitation \cite{Golez2024}.

Understanding X-ray absorption at an impurity site with an retarded (boson-mediated) interaction remains particularly challenging for time-resolved signals. A robust approach to solving quantum impurity models is the strong-coupling expansion, which systematically expands the interaction between the impurity and its environment \cite{Bickers1987, Coleman1984}. Its time-dependent variant \cite{Eckstein2010} is widely used in non-equilibrium DMFT simulations \cite{Aoki2014}. Bosonic modes can be incorporated by an expansion in the retarded interaction \cite{Golez2015} or using polaron representations \cite{Werner2013}. However, both methods have predominantly been applied in their lowest-order form (non-crossing approximation, NCA), and discrepancies between them highlight the need for systematic extensions \cite{Chen2016}.  In this paper, we aim to systematically investigate how higher-order corrections in the strong-coupling expansion for the retarded interaction can improve the description of the XAS signal in impurity models. To this end, we consider a simplified electron-boson model of the Holstein type \cite{Ament2011, Geondzhian2020, Hoffmann2002, Ament2011EPL}. This model is analytically tractable, but still captures the primary signatures of dynamic screening on XAS spectra such as peak shifts and sidebands. Simultaneously, it can serve as a benchmark for convergence of the strong-coupling expansion. 

To make the evaluation of high-order diagrams numerically feasible, we adapt the complex-mode decomposition method developed in \cite{Kaye2024} to the present real-time simulation. The key idea is to decompose the retarded interaction into a sum of complex exponential functions ("complex modes"), which simplifies the evaluation of strong-coupling diagrams.  By assuming a decomposition with $\chi$ modes, the computational cost of evaluating a $k$th-order diagram on a time axis with $N$ points scales as $\mathcal{O}(N \log N) \mathcal{O}(\chi^k)$ for $k \geq 2$. At nonzero temperatures, $\chi$ typically saturates with increasing $N$, making this approach significantly more efficient than direct equidistant time-grid evaluations, which scale as $\mathcal{O}(N^{2k-2})$ for \mbox{$k \geq 3$}. Using this algorithm, we can evaluate the strong-coupling expansion up to third order in the boson-induced interaction, which would otherwise be computationally prohibitive.

This paper is structured as follows. In Section \ref{sec:model}, we introduce the model and the calculation of the XAS signal within the strong-coupling expansion. In Sec.~\ref{sec:method}, we explain the complex-mode decomposition. Benchmarks for the latter  are given in Secs.~\ref{ssec:overview} to \ref{ssec:sigma}. Section \ref{ssec:res_time} and \ref{ssec:res_xas} show the convergence of the time-domain Green's function and the XAS line shapes, respectively. A conclusion and summary is provided in Sec.~\ref{sec:conclus}.

\section{Model}
\label{sec:model}

\subsection{Impurity model}
\label{ssec:XAS_model_H}

We consider a cluster with precisely one core level from which the X-ray absorption can happen.  The excitation site will be called the impurity site in the following.  The core excited electron is often tightly bound to the core hole due to a large electrostatic interaction, thus restricting the local electronic environment of the atom to a small cluster.  Here we entirely focus on the effect of  bosonic excitations on the XAS signal. Therefore, we take into account only a single electronic excited state in the cluster, but assume that the electron in this orbital interacts with charge fluctuations in the remaining lattice. This leads to an electron-boson model of the form
\begin{align}
	H_{d} = \epsilon_d n_d - \gamma  \sum_\alpha n_d (b^\dagger_\alpha+b_\alpha) + \sum_\alpha \omega_\alpha b^\dagger_\alpha b_\alpha,
	\label{eq:LF_Hamiltonian-1}
\end{align}
where $d_\sigma$ ($d_\sigma^\dagger$) represents the annihilation (creation) operator for an electron with spin $\sigma$ in the valence orbital  at energy $\epsilon_d$, and $n_d=\sum_\sigma d_\sigma^\dagger d_\sigma$ is the density; $b_\alpha$ ($b^\dagger_\alpha$) are annihilation (creation) operator for the bosonic modes. The latter represent the charge fluctuations, which interact via the (dimensionless) displacement operator $X_\alpha=(b^\dagger_\alpha+b_\alpha)/\sqrt{2}$ with the electron density. We will later take the continuum limit using the bosonic density of states
\begin{align}
\label{Abos}
	A_\text{b}(\omega) = \frac{1}{2} \sum_{\alpha}\big[\delta(\omega-\omega_\alpha) -\delta(\omega+\omega_\alpha)\big].
\end{align}
 The model \eqref{eq:LF_Hamiltonian-1} has been studied as an exactly solvable model to investigate the influence of bosonic fluctuations on the absorption process \cite{Ament2011, Geondzhian2020, Hoffmann2002, Ament2011EPL}. The analytical solution based on the Lang-Firsov transformation \cite{LangFirsov1962} is summarized in App.~\ref{app:ppgf_exact}. In the DMFT approach to XAS \cite{Cornaglia2007} the impurity model would also include a fermion environment, but here we focus on the description of the bosonic modes within the strong coupling expansion, for which the solvable model \eqref{eq:LF_Hamiltonian-1} provides an ideal benchmark. 
 
To the valence Hamiltonian we add a core level Hamiltonian $H_c=\sum_{\sigma} \epsilon_c c_\sigma^\dagger c_\sigma$, and the X-ray interaction
\begin{align}
H'&=
g\Big(i s(t) \mathrm e^{-i\omega_{\text{in}}t} P + h.c.\Big).
\end{align}
Here, $P=\sum_\sigma d^\dagger_\sigma c_\sigma$ is the dipolar transition operator for X-ray absorption (core hole creation), $s(t)$ is the envelope of the X-ray pulse and $\omega_{\text{in}}$ its energy; $g$ is an overall amplitude. The XAS  signal $I_\text{XAS}$ is defined as the photo-current at the detector due to the absorption of X-ray photons by the sample. Using standard response theory one obtains
\begin{align}
I_\text{XAS}= \lim_{g\to 0} \text{Re}\left( \frac{2}{g}\int_{-\infty}^\infty \mathrm dt\, s(t) \mathrm e^{-i\omega_\text{in} t} \langle P(t)\rangle_{g} \right),
\label{eq:Ixas}
\end{align}
where $\langle P(t)\rangle_{g}$ is the expectation value of the transition operator computed in the presence of the X-ray pulse (see Ref.~\cite{Werner2022} for a derivation using the Keldysh formalism). The limit $g\to0$ eliminates multi-photon absorption or emission (note that $\langle P(t)\rangle_{g}\sim g$).

\subsection{Action formulation}
\label{ssec:XAS_model_S}

In order to apply the strong coupling expansion, we integrate out the environment and represent the model by the action
$ S= S_{loc}+ S_{\Delta} $, where $S_{loc}$ is the action of the isolated site (core and valence orbitals),  and
\begin{align}
S_{\Delta} = -  \frac{1}{2} \int_{\mathcal C}\mathrm dt \mathrm dt' n_d (t)\Delta(t,t') n_d(t')
\end{align}
is the retarded interaction between the density $n_d = n_{d\uparrow}+ n_{d\downarrow}$  at different times. The action is defined such that $\langle \mathrm T_{\mathcal C} \dots \rangle = \text{tr} [\mathrm T_{\mathcal C} e^{iS} \cdots ]/\text{tr} [\mathrm T_{\mathcal C} e^{iS}]$ gives the contour-ordered averages. The interaction $\Delta$  is determined by the propagator of the bosonic  mode,
\begin{align}
	\Delta(t,t') = -2i\gamma^2 \sum_\alpha \big\langle
	\mathrm T_{\mathcal C}
	X_\alpha (t) X_\alpha (t')\big\rangle.
	\label{eq:Delta}
\end{align}
In terms of the spectral function \eqref{Abos} and the inverse temperature $\beta=1/k_{\rm B}T$, it is given by 
\begin{align}
\Delta^>(t-t') &= -2i\gamma^2 \int_{-\infty}^\infty \mathrm d\omega\,\mathrm e^{-i\omega(t-t')} \frac{A_\text{b}(\omega)}{1-\mathrm e^{-\beta\omega}},
\label{eq:Delta_gtr_time}
\end{align}
and $\Delta^<(t)=\Delta^>(-t)$.

\begin{figure}[tbp]
	\centerline{\includegraphics{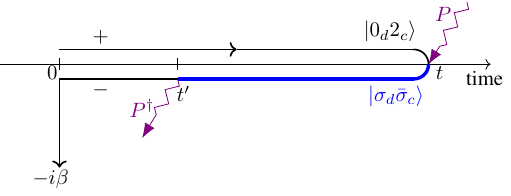}}
\caption{Keldysh contour with upper ($+$), lower ($-$), and imaginary branch, illustrating the XAS process with  the dipolar transition operator for X-ray absorption ($P$) and emission ($P^{\dagger}$); $\ket{0_d2_c}$ and $\ket{\sigma_d \bar \sigma_c}$ are the state before and after the absorption process, respectively.}
\label{fig:contour}
\end{figure}%

The leading contribution in $g$ to the expectation value $\langle P(t) \rangle_g$ in Eq.~\eqref{eq:Ixas} is then obtained by expanding $e^{iS}$ in the X-ray term  $S'=-\int dt H'$,
\begin{align}
\langle 
&
P(t) \rangle_g
=
\frac{-i}{Z_0}
\int_{\mathcal C}
\mathrm d t'\,
\text{tr}\Big[
\mathrm T_{\mathcal C}
e^{iS_{0}}
H'(t') P(t)
\Big]
\\
&=
-
\frac{g}{Z_0}
\int^0_{t}
\mathrm d t'\,
s(t')^* e^{i\omega_{\rm in} t'}
\text{tr}\Big[
\mathrm T_{\mathcal C}
e^{iS_{0}}
P^\dagger(t') P(t)
\Big],
\label{expval01}
\end{align}
where $S_0$ ($Z_0$) is the action (partition function) of the un-driven system ($g=0$). Terms of order $\mathcal{O}(g^2)$ have been omitted, and we have used that products  $ P(t')  P(t) $ give zero expectation values with the action $S_0$. The term in square brackets in Eq.~\eqref{expval01} is illustrated in  Fig.~\ref{fig:contour}, where black arrows represent evolution operators along the contour. The evolution along the vertical branch generates the initial density matrix $\sim e^{-\beta H}$. Because the core level is initially fully occupied, the operator $P^\dagger$ must be placed later than $P$ with respect to the contour ordering, i.e., within the interval $[0,t]$ on $\mathcal{C}_-$. The blue line segment marks the presence of the core hole. 

In our case, the valence orbital is empty before the X-ray absorption. Therefore, the boson-mediated interaction is only relevant while the core hole is present and the square bracket in Eq.~\eqref{expval01} can be written as $\sum_\sigma i\mathcal{G}_\sigma^>(t'-t) Z_0e^{-i2\epsilon_c(t-t')}$. The factor $Z_0e^{-i2\epsilon_c(t-t')} $ gives the contribution from the evolution of the system in the initial state $|0_d 2_c\rangle$ with both electrons in the core (black segments in Fig.~\ref{fig:contour}), and 
\begin{align}
\mathcal{G}_\sigma^>(t'-t) = -i \langle \sigma_d\bar\sigma_c | \mathrm T_{\mathcal C} e^{iS(t'\succ t)} | \sigma_d\bar\sigma_c \rangle 
\label{ppG1}
\end{align}
is related to the evolution operator of the state where spin $\sigma$ is excited to the valence orbital (blue segment in Fig.~\ref{fig:contour}). The action $S(t'\succ t)$ is restricted to the interval $(t',t)$ on $\mathcal{C}$ with $t'$ later than $t$ along $\mathcal{C}$. Because of spin symmetry, we can use $\mathcal{G}_\sigma=\mathcal{G}_{\bar\sigma}\equiv \mathcal{G}_1$, and $\sum_\sigma\mathcal{G}_\sigma =2\mathcal{G}_1$. Moreover, it is convenient to separate out from $\mathcal{G}_1$ the evolution operator of the unscreened impurity site (no bosons) with energy $\epsilon_d+\epsilon_c$,
\begin{align}
\mathcal{G}_1^>(t'-t) &\equiv \tilde{\mathcal{G}}_1^>(t'-t) \, e^{-i(\epsilon_d+\epsilon_c)(t'-t)} 
\nonumber \\
	&= i \tilde{\mathcal{G}_1}^>(t'-t) \, \mathcal G_{1,\rm{bare}}(t'-t);
\label{ppG1-1}
\end{align}
$ \tilde{\mathcal{G}}_1$ now contains all corrections due to screening. Inserting this expression into Eq.~\eqref{expval01}, and Eq.~\eqref{expval01} into Eq.~\eqref{eq:Ixas}, yields
\begin{align}
I_\text{XAS}
&= -4 \,\text{Im} \int \mathrm dt\,
\int_{-\infty}^{t} \mathrm dt' s(t) s(t')^* e^{i(\omega_{\rm in}-E_{\rm ex})(t'-t)}
\nonumber
\\
&\times\,\,\,
\tilde{\mathcal{G}}_1^>(t'-t),
\label{eq:Ixas_ret}
\end{align}
where $E_{\rm ex}=\epsilon_d-\epsilon_c$ is the bare X-ray transition energy. Using the hermitian property $ \mathcal{G}_1^>(t)= -\mathcal{G}_1^>(-t)^*$ of the operator \eqref{ppG1}, the integrals are extended over the full time domain,
\begin{align}
I_\text{XAS}
&= -2 \int \mathrm dt \mathrm dt' s(t)s(t')^* e^{i(\omega_{\rm in}-E_{\rm ex})(t'-t)} \tilde{\mathcal{G}}_1^>(t'-t).
\nonumber 
\end{align}
For an infinite probe pulse, $s(t) \to 1$, we can then extract the XAS rate (absorbed number of photons per unit time),
\begin{align}
\iota_\text{XAS}
&=
-2 \int \mathrm dt \,e^{i(\omega_{\rm in}-E_{\rm ex})t} \tilde{\mathcal{G}}_1^>(t).
\\
&=
-4\,\text{Im}\, \int_0^\infty \mathrm dt \,e^{i(\omega_{\rm in}-E_{\rm ex})t} \tilde{\mathcal{G}}_1^>(t).
\label{eq:ixas-nogamma}
\end{align}

Due to decay processes such as Auger-Meitner-decay the lifetime of the core hole is typically only of the order of a few femtoseconds \cite{Ament2011}. Because various decay channels contribute, it is often sufficient to model the decay by a heuristic, overall core-hole lifetime $1/\Gamma$, corresponding to a modification of the core excited propagator \eqref{ppG1} into 
\begin{align}
 \tilde{\mathcal{G}}_1^>(t'-t) \to   \tilde{\mathcal{G}}_1^>(t'-t)e^{-\Gamma |t-t'|}.
\end{align}
(This decay could also be implemented by coupling the core level to a completely filled fermion reservoir  \cite{Werner2022}.) With this we obtain the final expression for the XAS rate,
\begin{align}
\iota_\text{XAS}
&=
-4\,\text{Im}\, \int_0^\infty \mathrm dt \,e^{i(\omega_{\rm in}-E_{\rm ex} +i\Gamma)t}\, \tilde{\mathcal{G}}_1^>(t).
\label{eq:ixas}
\end{align}
The remaining task is to calculate the evolution operator \eqref{ppG1}. The exact solution (see Appendix \ref{app:ppgf_exact}) is given by 
\begin{align}
\mathcal{G}^>_{1}(t) = -i e^{-i(\tilde\epsilon+\epsilon_c)t}e^{f^>(t)},
\label{eq:G_exact}
\end{align}
where $f^>(t)$ is specified in Eq.~\eqref{eq:f>}, and 
\begin{align}
\tilde\epsilon=\epsilon_d - \gamma^2 \int_0^\infty \mathrm d \omega \, 2 A_{\rm b}(\omega)/\omega,
 \label{eq:redshift}
\end{align}
is the red-shifted energy level. In the next section, we explain how to compute the propagator \eqref{ppG1} using the strong-coupling expansion.  

\subsection{Strong-coupling expansion}
\label{ssec:strongcoupling}

\begin{figure}[t]
\begin{minipage}{\linewidth}
	\centerline{ \includegraphics{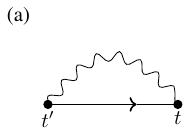} \hfill \includegraphics{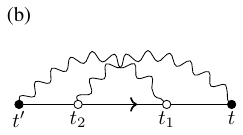} }
\end{minipage} \\
	\centerline{\includegraphics{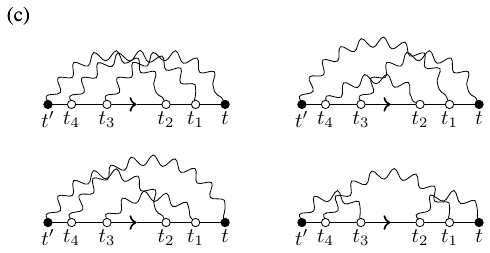}}
\caption{Self-energy $\Sigma^>(t-t')$ at (a) first order (NCA), (b) second order (OCA) and (c) third order in the interaction $\Delta^>$ \eqref{eq:Delta_gtr_time}. Solid lines from time $t_{j+1}$ to $t_j$ represent $ \mathcal{G}_1^>(t_{j}-t_{j+1})$, dots represent the interaction vertex $n_d$ (which is unity for the given model), and wiggly lines  represent the interaction $\Delta^>(t_i-t_j)$  ($t_i \succ t_j$). }
\label{fig:self-energy-diags}
\end{figure}

The strong-coupling expansion is a self-consistent diagrammatic expansion for the resolvent operators $\mathcal{G}$   (also called  pseudo-particle propagators) in terms of the non-local part $S_\Delta$ in the action.  Because the exact solution for $\mathcal{G}$ has a complex functional dependence on $\Delta$, the  model \eqref{eq:LF_Hamiltonian-1} constitutes a nontrivial benchmark for the strong-coupling expansion.  The derivation of the strong-coupling expansion on the Keldysh contour is detailed, e.g., in \cite{Eckstein2010,Aoki2014}. Here we will first recapitulate the final equations, and then focus on the discussion of their numerical solution.

Corrections to $\mathcal{G}_1^>(t)$ with respect to the bare evolution $\mathcal{G}_{1,\text{bare}}^>(t)=-ie^{-i(\epsilon_d+\epsilon_c)t}$ are constructed in terms of a self energy $\Sigma^>(t)$, and incorporated into $\mathcal{G}_1^>(t)$ by solving a Dyson equation. For $t>0$, the latter reads
\begin{align}
(i\partial_t -\epsilon_d-\epsilon_c) \mathcal{G}^>_1(t) - \int_{0}^t \mathrm d\overline t\, \Sigma^>(t-\bar t)  \mathcal{G}^>_1(\bar t)=0,
\label{eq:dyson_ret}
\end{align}
with initial condition $\mathcal{G}^>(0)=-i$.  The $m$th order self-energy is obtained by a sum of all possible diagrams consisting of a single sequence of $2m-1$ propagators $ \mathcal{G}_1^>$, whose endpoints are connected by $m$ interaction lines $\Delta$. All diagrams up to third order are shown in Fig.~\ref{fig:self-energy-diags}.
To evaluate a diagram, one integrates over the intermediate times while retaining their respective order: For the propagator $ \mathcal{G}_1^>$, all vertices are on the same branch of $\mathcal{C}$ (such as the blue segment in Fig.~\ref{fig:contour}), and the time order of the integration variables implies $t'\prec t_{2m-2}\prec \ldots  \prec t_1 \prec t$ if $t'\prec t$ (and $t'\succ t_{2m-2}\succ\ldots \succ t_1 \succ t$ if $t'\succ t$). The wiggly line in Fig.~\ref{fig:self-energy-diags} connecting two vertices $t_i$ and $t_j$ represents the boson-mediated interaction $\Delta(t_i,t_j)=\Delta^{>(<)}(t_i-t_j)$ if $t_i$ is later (earlier) on $\mathcal{C}$ than $t_j$.
Finally, a  diagram of order $m$ has an overall factor $i^m$.  For example, for the greater Keldysh part of the first (non-crossing approximation, NCA) and second order contribution (one-crossing approximation, OCA), one finds
\begin{align}
	\Sigma^{>}_{\text{NCA}}(t) =&\, i \,\Delta^>(t)\mathcal{G}_1^>(t) ,
	\label{eq:Sigma_NCA}
	\\
	\Sigma^{>}_{\text{OCA}}(t) =& \,i^2 \int_{0}^{t} \mathrm dt_1 \int_{0}^{t_1} \mathrm dt_2 \, \Delta^>(t-t_2)  \Delta^>(t_1)
	\nonumber\\
		& \quad  \times   \mathcal{G}_1^>(t-t_1) \mathcal{G}_1^>(t_1-t_2)  \mathcal{G}_1^>(t_2).
	\label{eq:Sigma_OCA}
\end{align}
Since the self energy is a functional of $\mathcal{G}_1$, Eq.~\eqref{eq:dyson_ret} has to be solved iteratively until converging to a self-consistent solution. While the solution of the Dyson equation for a given $\Sigma^>$ is numerically straightforward (we use the retarded Dyson equation solver of the NESSi library \cite{nessi2020}), the evaluation of the self-energy diagrams presents a considerable challenge due to the high order integration. For this purpose we introduce an efficient algorithm based on the decomposition of $\Delta$ into complex mode functions in the next section.

\section{Real-time complex mode decomposition}
\label{sec:method}

\begin{figure}[t]
\centering
	\centerline{\includegraphics[width=0.8\columnwidth]{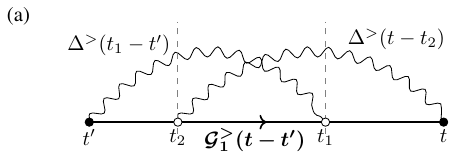}} 
	\centerline{\includegraphics[width=0.8\columnwidth]{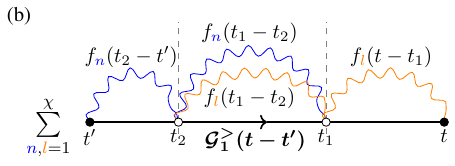}}

	\caption{OCA self-energy $\Sigma^>_{\rm OCA}(t-t')$, in terms of (a) interaction $\Delta^>$ and (b), after decomposing $\Delta^>$ according to Eq.~\eqref{eq:hybrid_decomp}. 
	}
	\label{fig:feynmandiagram}
\end{figure}

We now focus on the numerical evaluation of the self-energy diagrams beyond first order, extending the formalism developed in Ref.~\cite{Kaye2024} from imaginary time to a single real-time contour. The main concept can be explained for the first nontrivial diagram (OCA) \eqref{eq:Sigma_OCA}.  
We assume that the interaction can be approximated as a sum of $\chi$ complex exponentials, 
\begin{align}
\label{ansatz}
\Delta^>(t-t') &= \sum_{n=1}^\chi b_n z_n^{t-t'}, \,\,\,\text{for~~}(t-t')>0,
\end{align}
with given parameters $b_n$ and $z_n$ ($|z_n|<1$). Various algorithms exist to obtain such a decomposition starting from a given function $\Delta^>(t)$ on an equidistant grid. A well known algorithm is Prony's method, dating back to 1795 \cite{Prony1795}, and its extensions  \cite{Peter2013,Sauer2017}, which are applied to a wide range of signal processing problems. An alternative is the generalized matrix pencil approach \cite{Sarkar1995}, which is more robust for noisy data and can often  obtain more compact representations even for noiseless data. For the purpose of the present work, we use a variant of Prony's method, which is summarized in App.~\ref{ssec:cplx_mode_decomp}.
With the ansatz \eqref{ansatz}, $\Delta^>$ can be brought to the form
\begin{align}
\label{eq:hybrid_decomp}
\Delta^>(t-t') &= \sum_{n=1}^\chi b_n f_n(t-\overline t) f_n(\overline t -t'),
\end{align}
for any intermediate time $\overline t$ with $t\geq \overline t \geq t'$, where  the functions $f_n(t)=z_n^t$ do not depend on $\overline t$. If the splitting points $\bar t$ are chosen to be another vertex time [see dashed lines in~Fig.~\ref{fig:feynmandiagram}(a)],  the OCA diagram separates into $\chi^2$ terms, each containing four instead of two interaction lines, see Fig.~\ref{fig:feynmandiagram}(b). The integral expression for $\Sigma^{>}_{\text{OCA}}$ becomes
\begin{align}
&\Sigma^{>}_{\text{OCA}}(t) = \,i^2 \int_{0}^{t} \!\!\mathrm dt_1 \int_{0}^{t_1} \mathrm dt_2 \,   
\mathcal{G}_1^>(t-t_1) \mathcal{G}_1^>(t_1-t_2)  \mathcal{G}_1^>(t_2)
\nonumber\\
&\times
\sum_{n,l=1}^\chi  b_{n}b_l f_l(t-t_1) f_l(t_1 -t_2) f_n(t_1-t_2) f_n(t_2).
\label{eq:Sigma_OCA_example_decomp}
\end{align}
We may rewrite the integral in a more compact form
\begin{align}
\Sigma^{>}_{\text{OCA}}(t) = &
i^2
\sum_{n,l}b_n b_l
\, \int_{0}^{t} \!\!\mathrm dt_1 \int_{0}^{t_1} \mathrm dt_2 \,   
B^{(1)}_{l}(t-t_1)
\,\,\,
\nonumber\\
&\,\,\,\times\,\,\,\, B^{(2)}_{ln}(t_1-t_2)  B^{(3)}_{n}(t_2),
\label{eq:Sigma_OCA_example_decomp-A}
\end{align}
with
\begin{align}
B^{(1)}_{l}(t)&=\mathcal{G}_1^>(t)f_l(t),
\\
B^{(2)}_{ln}(t)&=\mathcal{G}_1^>(t)f_l(t)f_n(t),
\\
B^{(3)}_{n}(t)&=\mathcal{G}_1^>(t)f_n(t).
\end{align}

This shows the main advantage of the decomposition: Given $n,l$, the integrals can now be calculated successively, as
\begin{align}
C_{ln}^{(1)}(t_1)
&= \int_{0}^{t_1} \mathrm dt_2  B^{(2)}_{ln}(t_1-t_2)  B^{(3)}_{n}(t_2),
\label{eq:conv1}
\\
\Sigma^{>}_{\text{OCA},ln}(t) 
&= 
i^2 \, \int_{0}^{t} \!\!\mathrm dt_1 
B^{(1)}_{l}(t-t_1) C_{ln}^{(1)}(t_1),
\label{eq:conv2}
\\
\Sigma^{>}_{\text{OCA}}(t) 
&= \sum_{l,n=1}^\chi b_n b_l\, \Sigma^{>}_{\text{OCA},ln}(t) .
\label{eq:Sigma_OCA_example_decomp-B}
\end{align}
For time $t$ taking $N$ possible values, the convolution integrals \eqref{eq:conv1} and \eqref{eq:conv2} can be solved for all $t$ using a fast Fourier transform at an effort $\mathcal O(N \log N)$. The number of terms in Eq.~\eqref{eq:Sigma_OCA_example_decomp-B} in turn increases by a factor of $\chi^2$, such that the total effort is $\mathcal O(\chi^2 N\log N)$. In comparison, the effort for a direct evaluation of the OCA diagrams is $\mathcal{O}(N^3)$, as a double integral \eqref{eq:Sigma_OCA} is required for each external time point $t$. Hence the integration using the decomposition will be more efficient if the number of modes  $\chi$ required in the decomposition is $\chi\ll N$. Moreover, the sums over the decomposition indices $n$ and $l$ can be parallelized easily.

The scheme can be readily generalized to higher orders. At order $m$ there are $m$ interaction lines to be decomposed. For each set of decomposition indices, the integrals again reduce to a set of convolutions, such that the effort scales like $\mathcal O(\chi^m N\log N)$. Instead, direct integration with one external argument and $(2m-2)$ integrals corresponds to an effort $\mathcal O(N^{2m-1})$; by factoring out and pre-calculating  a triangular vertex (similar to \cite{AramKim2023}), the complexity can be reduced to $\mathcal O(N^{2m-2})$ for $m\ge 3$. Therefore, even if $\chi$  would scale linearly in $N$, the decomposition-based integration is expected to be advantageous for orders $m>3$, and for all orders we expect a speedup if $\chi \ll N$.

We remark that the summation over the decomposition indices can be done more efficiently than $\mathcal{O}(\chi^m)$ if the diagram has a simpler structure. For example, the fourth diagram for order $m=3$ in Fig.~\ref{fig:self-energy-diags}(c) can be written as 
\begin{align}
&i^3\sum_{n,r,l} b_n b_r b_l
 \int_0^t \mathrm dt_1
\int_0^{t_1} \mathrm dt_2
\int_0^{t_2} \mathrm dt_3
\int_0^{t_3} \mathrm dt_4 \,B^{(1)}_n(t-t_1)
\nonumber\\&
\times\,\,\,
B^{(2)}_{nr}(t_1-t_2)
B^{(3)}_{rr}(t_2-t_3)
B^{(4)}_{rl}(t_3-t_4)
B^{(5)}_l(t_4),
\nonumber
\end{align}
with 
$B^{(1)}_{n}(t)=\mathcal{G}_1^>(t)f_n(t),
$ $
B^{(2)}_{n,r}(t)=\mathcal{G}_1^>(t)f_r(t)f_n(t),
$ $
B^{(3)}_{r,r}(t)=\mathcal{G}_1^>(t)f_r(t),
$ $
B^{(4)}_{r,l}(t)=\mathcal{G}_1^>(t)f_l(t)f_r(t),
$ $
B^{(5)}_{l}(t)=\mathcal{G}_1^>(t)f_l(t).
$
Hence, the summation over $l,n,r$ can be performed as a successive matrix vector product in only $\mathcal{O}(\chi^2)$ operations.  However, at each order there should remain at least one  sufficiently connected diagram such that the summation over the decomposition indices cannot be reduced below $\mathcal{O}(\chi^m)$.

\begin{figure}[tbp]
  \begin{minipage}[b]{0.25\textwidth}
  		\raggedright (a) \\
     \centerline{\includegraphics[width=\textwidth]{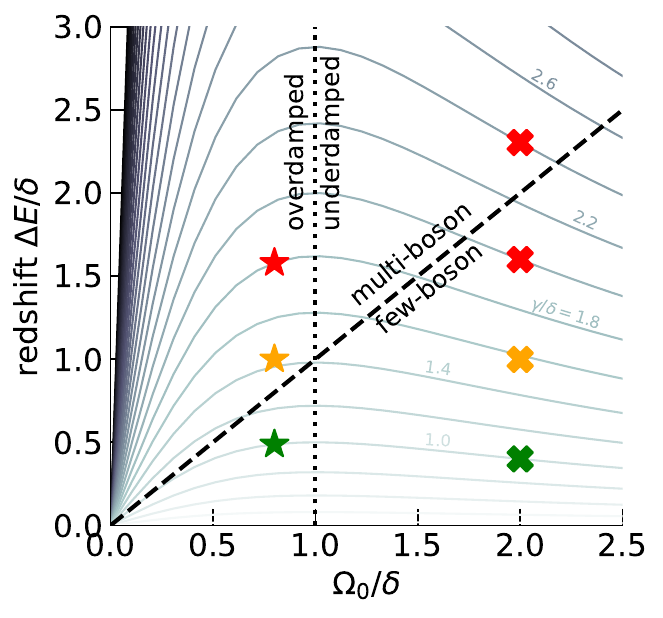}}
  \end{minipage} \hfill
   \begin{minipage}[b]{0.22\textwidth}
   		\raggedright (b) \\
     \centerline{\includegraphics[width=\textwidth]{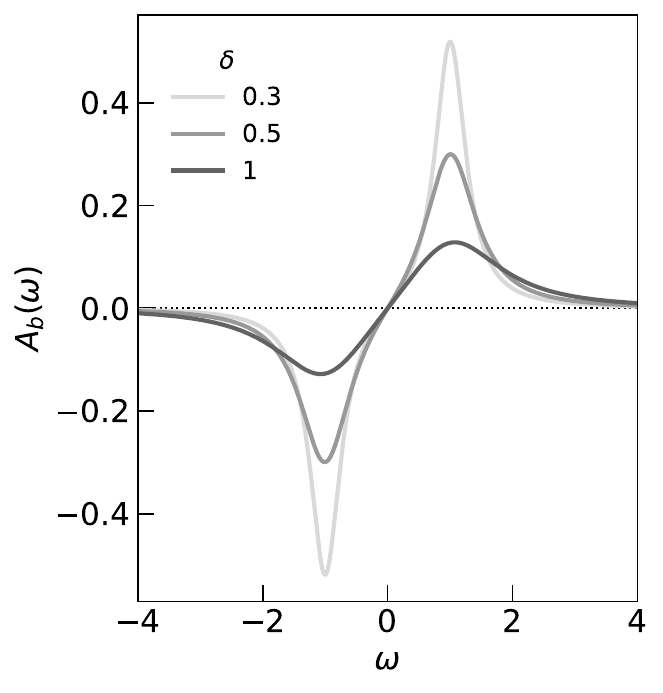}}
  \end{minipage} \hfill
\caption{(a) Different parameter regimes of the model. The markers indicate the parameter sets for which XAS signals are shown in Fig.~\ref{fig:results_XAS}. The thin lines correspond to a fixed value of $\gamma/\delta$. (b) Spectral function $A_b(\omega)$ [Eq.~\eqref{eq:A_Delta}] for $\Omega_0=1$ and varying damping $\delta$.}
\label{fig:overview}
\end{figure}

\section{Results}
\label{sec:results}

\subsection{Overview}
\label{ssec:overview}

\begin{figure}[t!]
	(a) \raggedright
	\centerline{\includegraphics[width=0.9\linewidth]{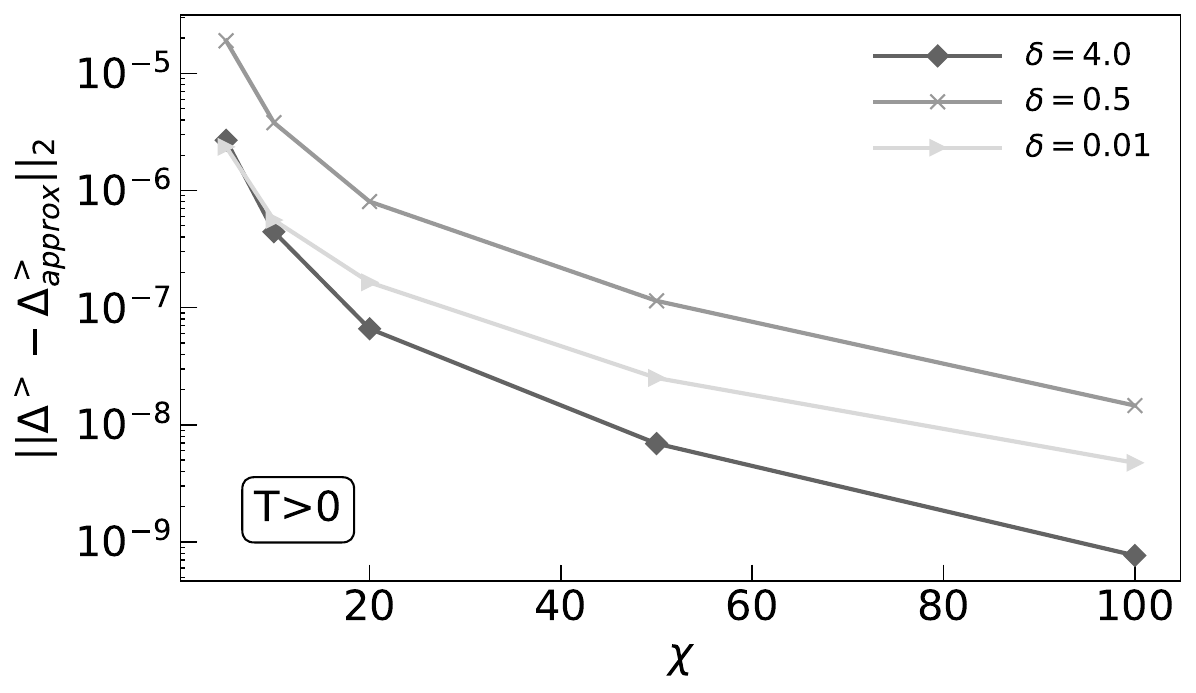}}
	(b)
	\centerline{\includegraphics[width=0.9\linewidth]{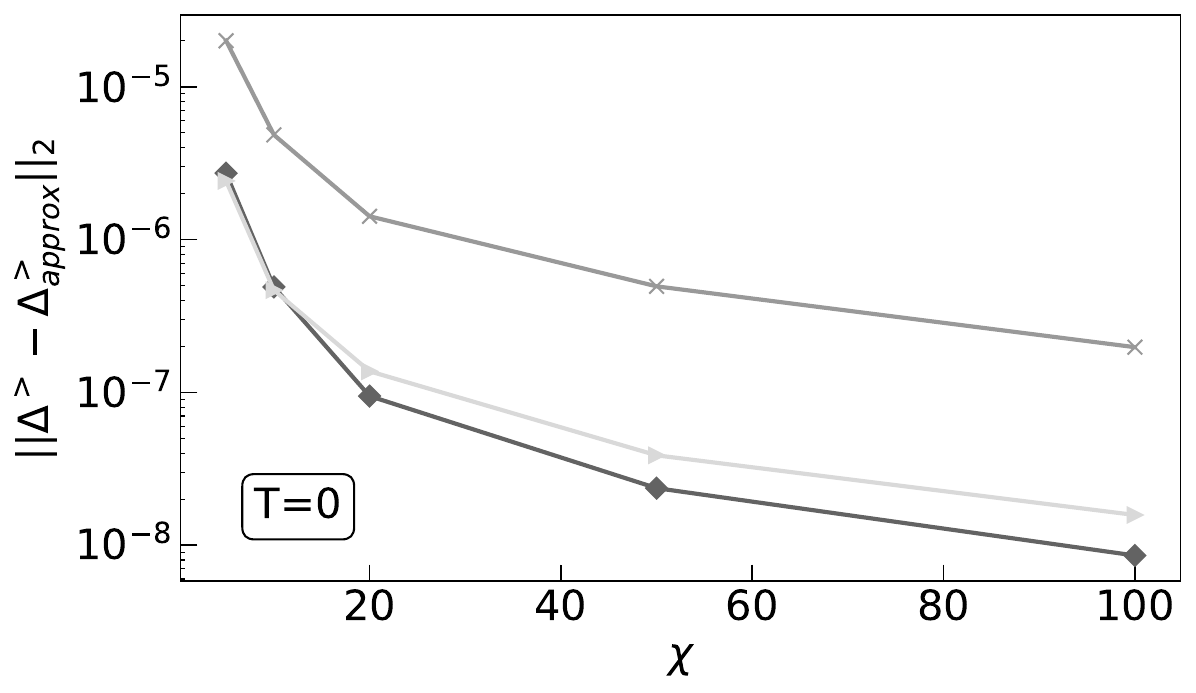}}
\caption{Deviation of the decomposed interaction \eqref{eq:hybrid_decomp} from the exact expression \eqref{eq:Delta_gtr_time}, as a function of the number of modes $\chi$ at temperature $T=1/50$ (a) and $T=0$ (b), 
for the strongly damped regime ($\delta=4$), the weakly damped regime ($\delta=0.5$) and the almost undamped regime ($\delta=0.01$). Other parameters are $\Omega_0=1,\gamma=1, \delta t=0.02,N=8193$.}
\label{fig:decomp_examples}
\end{figure}~
\begin{figure*}[tpb]
	\begin{minipage}[b]{0.42\textwidth}
		\raggedright (a)		\\
		\centerline{		\includegraphics[width=\textwidth]{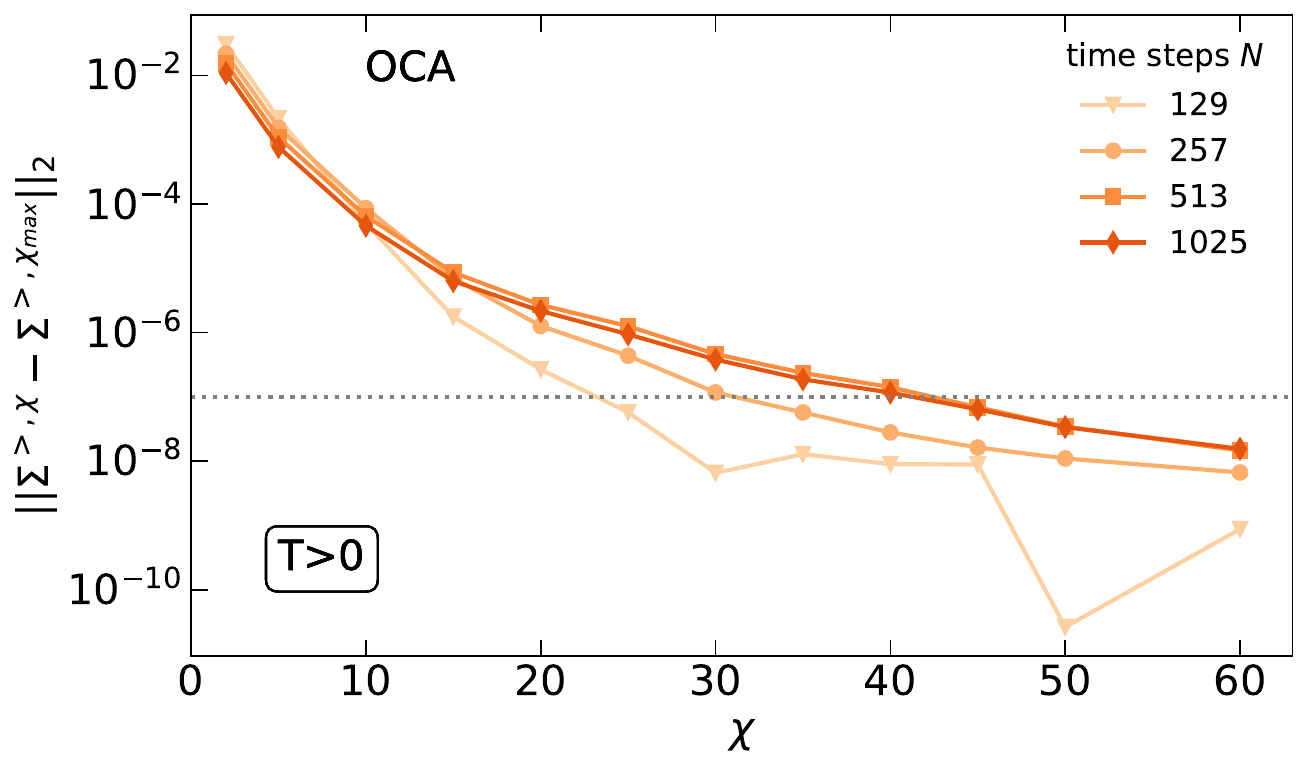}  }
	\end{minipage} ~~~~~
	\begin{minipage}[b]{0.42\textwidth}
		\raggedright (c) 		\\
		\centerline{ 		\includegraphics[width=\textwidth]{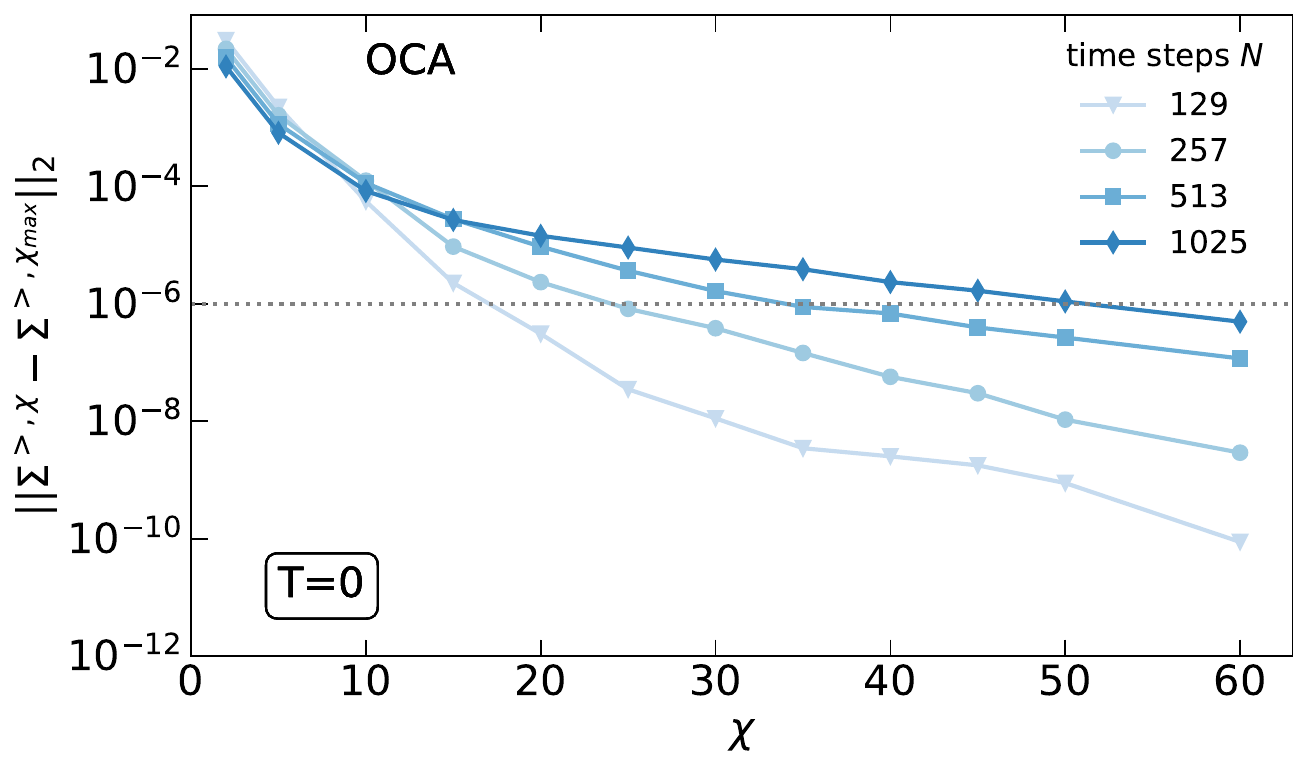}	}
	\end{minipage} \\
		\begin{minipage}[b]{0.42\textwidth}
		\raggedright (b)		\\
		\centerline{	\includegraphics[width=\textwidth]{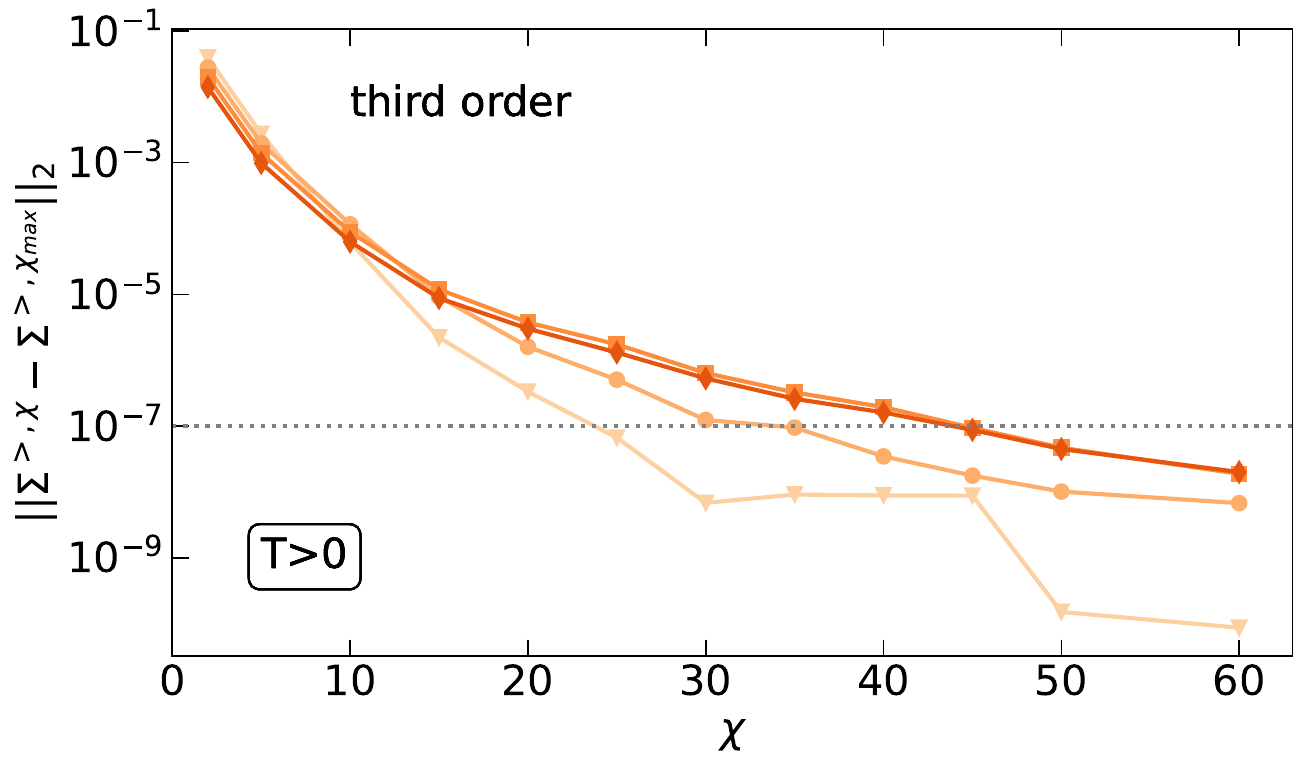}   }
	\end{minipage} ~~~~~
	\begin{minipage}[b]{0.42\textwidth}
		\raggedright (d)       \\		
		\centerline{ 	\includegraphics[width=\textwidth]{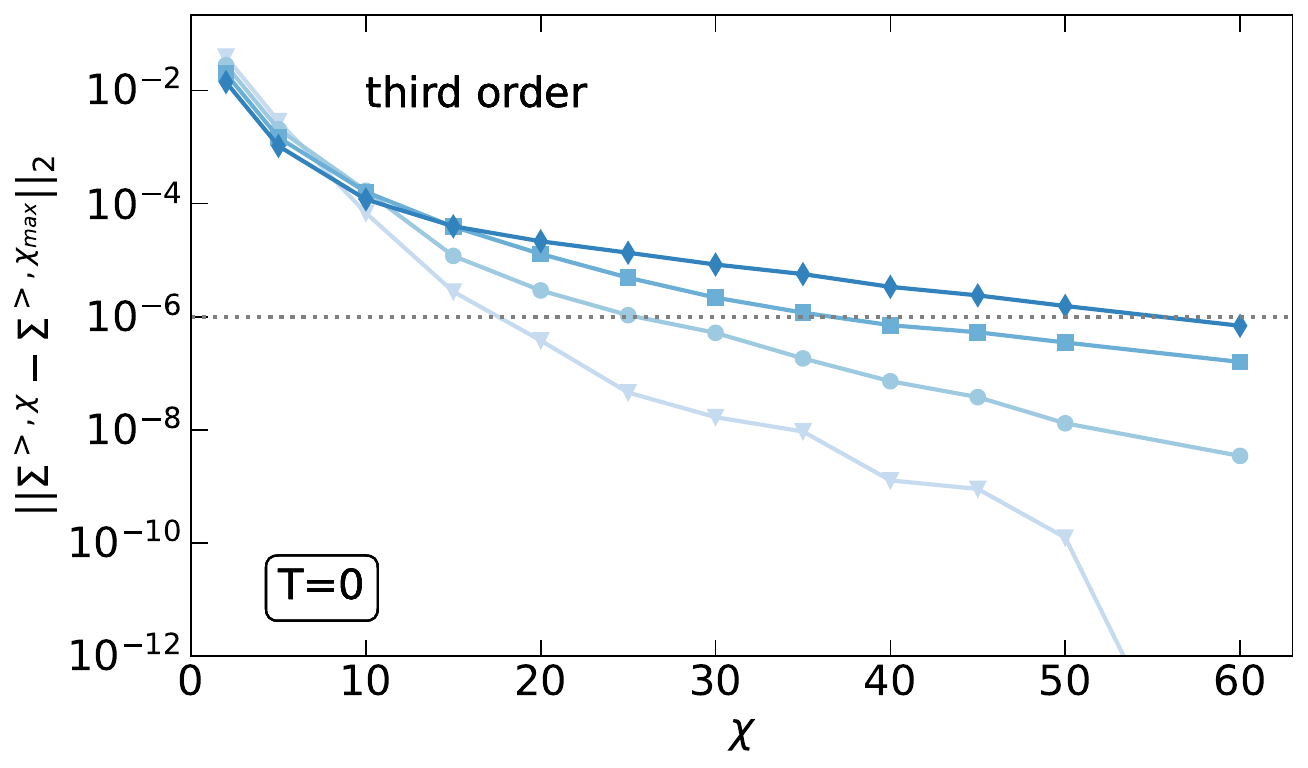}	}
	\end{minipage} 
	\caption{Deviation of the bare OCA [upper plots (a,c)] and third order [lower plots (b,d)] self energy $\Sigma^>$, calculated with a complex mode decomposition of the hybridization function with $\chi$ modes, from the same utilizing $\chi_{\text{max}}=100$ decomposition modes. Other parameters are $\delta=1, \Omega_0=4,\gamma=2$, the time step size is $\delta t = 0.1$ and the temperature $T=1/50$ [left plots (a,b)] and $T=0$ [right plots (c,d)].}
	\label{fig:conv_test}
\end{figure*}

For the numerical analysis of the model \eqref{eq:LF_Hamiltonian-1} we will choose the spectrum
\begin{align}
	A_\text{b}(\omega) &= \frac{1}{4\pi}\left(\frac{2\delta}{(\omega-\Omega_0)^2+\delta^2} - \frac{2\delta}{(\omega+\Omega_0)^2+\delta^2} \right),
		\label{eq:A_Delta}	
\end{align}
with a linear onset  $A_\text{b}(\omega) \sim\omega$ for small $\omega$, and a Lorentzian peak of width $\delta$ around $\omega=\pm\Omega_0$, see Fig.~\ref{fig:overview}(b); $A_\text{b}(\omega)$ corresponds to the spectrum of a damped harmonic oscillator. Figure~\ref{fig:overview}(a) gives an overview over different relevant parameter regimes of the model: (i) If $\Omega_0<\delta$, the screening mode is weakly damped and $A_b(\omega)$ exhibits a sharp peak at $\omega=\Omega_0$, while $\Omega_0\gtrsim \delta$ marks the strongly damped or overdamped regime. (ii) The coupling strength $\gamma$ can be quantified by the redshift $\Delta E=\epsilon_d-\tilde \epsilon$ of the main absorption peak [see Eq.~\eqref{eq:redshift}], which becomes $\gamma^2/\Omega_0$ in the weakly damped regime. For $\Omega_0 > \Delta E$, a charge excitation on the impurity, as created by the X-ray pulse, typically creates only few excitation quanta in the screening environment (few-boson regime). Conversely, the multi-boson regime is defined by $\Omega_0 < \Delta E$ and the XAS line shape will exhibit more features than a simple redshift, such as~broadening and asymmetry. Note that for a single harmonic screening mode, as in \cite{Golez2024}, the distinction between a few- and many-boson regime is equivalent to separating into an adiabatic ($\gamma/\Omega_0>1$) and an anti-adiabatic ($\gamma/\Omega_0<1$) regime.

\subsection{Complex mode decomposition for the hybridization}
\label{ssec:decomp_applied}

First, we demonstrate the accuracy of the complex mode decomposition for a given interaction $\Delta^>(t)$, for  $\Omega_0=\gamma=1$ and various values of the damping $\delta$. An approximation $\Delta^>_{\rm approx}(t)$ given by Eq.~\eqref{eq:hybrid_decomp} with $\chi$ complex modes is constructed from the exact function $\Delta^>(t)$ on an equidistant grid ($N=8193$ points with timestep $\delta t=0.02$). In Fig.~\ref{fig:decomp_examples}, we analyze the reconstruction error $||\Delta^> - \Delta^>_{\rm approx}||_2$, given by the $2$-norm difference between the functions on the time grid.

In particular, it is interesting to compare the decomposition at nonzero and zero temperature $T$. For $T>0$, the integrand in Eq.~\eqref{eq:Delta_gtr_time} is an analytic function, hence $\Delta^>(t)$ decays exponentially at large times. On the other hand, at $T=0$, $\Delta^>(t)$ exhibits an algebraic decay $\sim 1/t^2$, as the integrand is no longer differentiable at $\omega=0$. Fitting the algebraic tail with complex exponentials is more challenging, which is reflected in a slower decrease of the approximation error with respect to the number of modes $\chi$. 
(At $\chi=100$, the algebraically decaying hybridization $\Delta^> (t)$ reduces the fidelity of the complex mode decomposition by about one order of magnitude.) 
Nevertheless, in either case, one can achieve an accurate representation using a number of complex modes that is sufficiently small to facilitate the evaluation of third order diagrams. 

We note that the complex mode decomposition performs better for the strongly damped case ($\delta=4$) that for weak damping ($\delta = 0.5$). This is simply because for large $\delta$, both the exact $\Delta^>$ and the reconstructed $\Delta^>_{\rm approx}$ decay faster, automatically leading to a smaller integrated error. Finally, the error decreases naturally for $\delta\to0$. In this case, $A_b(\omega)$ approaches a sum of two $\delta$-functions such that $\Delta^>(t)$ is given by exactly a sum of two complex exponentials [cf.~Eq.~\eqref{eq:Delta_gtr_time}].

\subsection{Accuracy of the self energy integrals}
\label{ssec:sigma}

In order to evaluate the efficiency of the complex mode decomposition, we investigate the error in the self energy $\Sigma^>$ with the time steps $N$ and the number of modes $\chi$. For the benchmark in this subsection, the integral for $\Sigma^>$ is evaluated with the the non-interacting pseudo-particle Green's  function $\mathcal G^>_{1,\rm bare}$  as given in Sec.~\ref{ssec:XAS_model_S}. As error measure we take the 2-norm difference between the respective self energy for given $\chi$ and the one for $\chi_{\text{max}}=100$ decomposition modes.

Figures \ref{fig:conv_test}(a,b) show the deviation of the finite-temperature OCA (a) and third order (b) self energy with $\chi$ modes and $N$ time steps. Both the error in the OCA and third order diagrams decays with $\chi$.  For the third order, the deviation is larger, consistent with the fact that its absolute value decays more slowly in time as compared to the OCA contribution. Furthermore, the error in the self energy becomes larger the more time steps are included, but it saturates for $N\geq 513$.  For larger $N$,  the exponentially decaying tail in the hybridization $\Delta^>(t)$ has decayed sufficiently within the integration interval, which is automatically accounted for by the complex mode decomposition. For zero temperature, the same analysis is presented in Fig.~\ref{fig:conv_test}(c,d). In this case, we cannot observe a saturation of the error with the number of time steps at fixed $\chi$, because the algebraic tails of $\Delta^>(t)$ give a significant contribution to the convolution integrals up to the maximum time considered here. Nevertheless, the integration based on the complex mode decomposition is far more efficient than a direct integration, as the following analysis will demonstrate.

\begin{figure}
\centerline{\includegraphics[width=0.9\linewidth]{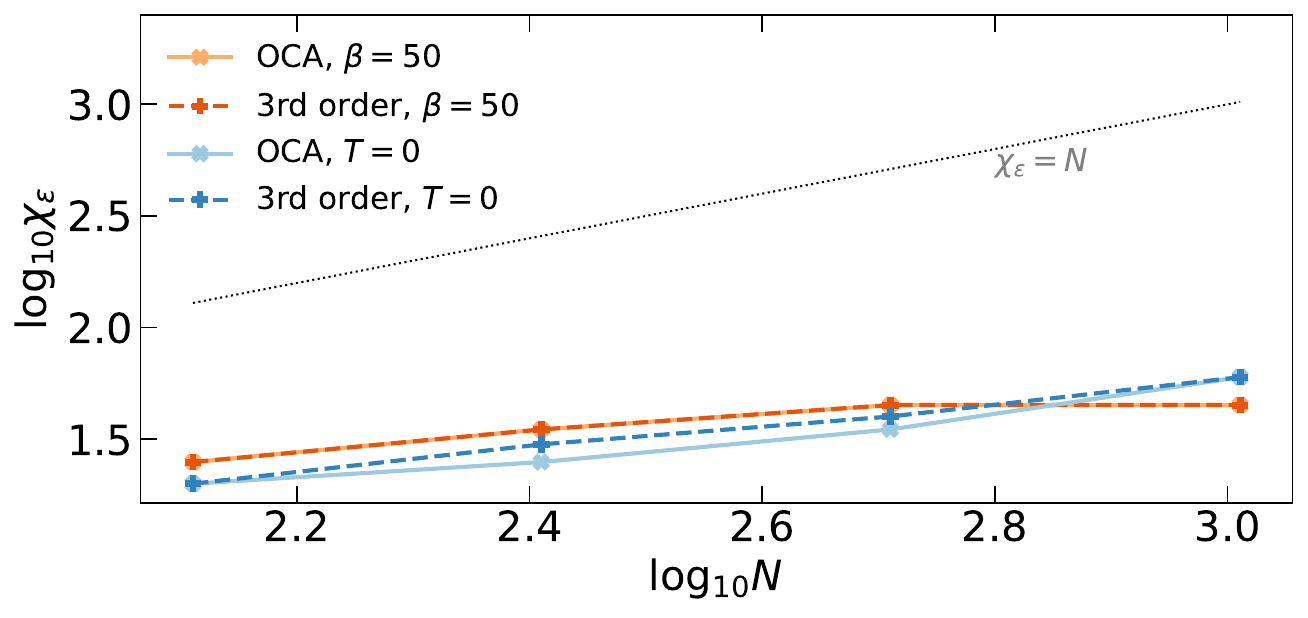}}
\caption{Number of modes $\chi_\epsilon$ needed to achieve a deviation in the self energy of at most $10^{-7}$ ($T=1/50$) and $10^{-6}$ ($T=0$) over a time interval of $N$ steps; see dotted line in Figs.~\ref{fig:conv_test}(a-d) for the error thresholds. Other parameters are the same as in Fig.~\ref{fig:conv_test}.}
\label{fig:chi_vs_N}
\end{figure}

\begin{figure}[tbp]
	(a) \raggedright
	\centerline{\includegraphics[width=\linewidth]{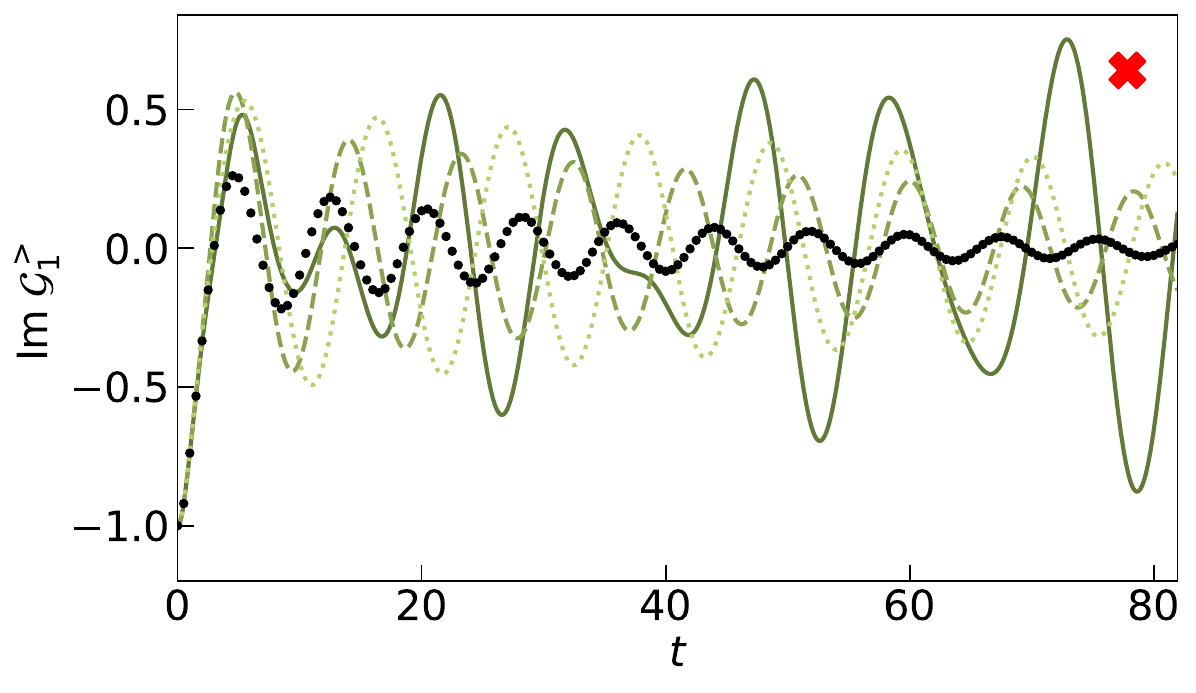}}
	(b) \raggedright
	\centerline{\includegraphics[width=\linewidth]{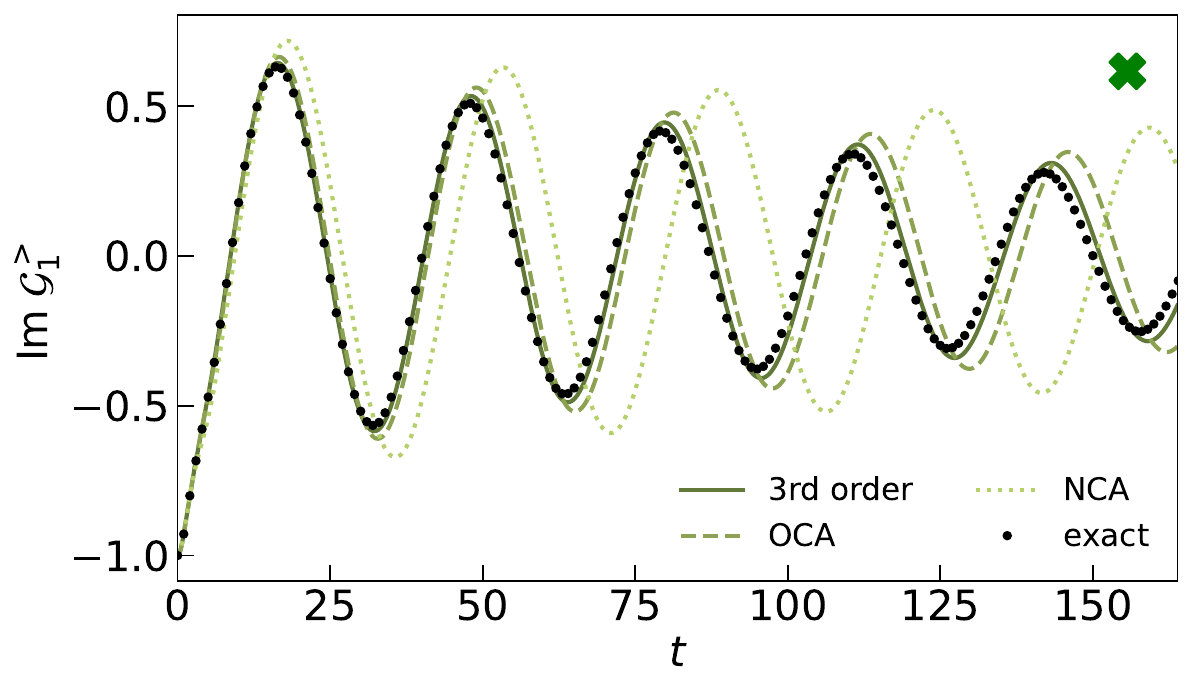}}
	\caption{Imaginary part of the Green's function $\mathcal G^>_1(t)$ 
	including all diagrams up to first order (NCA, light green dotted), second order (OCA, green dashed) and third order (dark green, solid line) in the self energy $\Sigma^>$. We set $\epsilon_d + \epsilon_c =0$,  the temperature is $T=1/50$, and $\chi=50$ modes were used in the decomposition. Black dots show the analytic result \eqref{eq:G_exact}. (a) shows data  in the vicinity of the crossover between multi-boson and few-boson regime ($\Delta E = 0.8, \Omega_0=1,\delta =0.5$). (b) is deep in the few-boson regime ($\Delta E=0.25, \Omega_0=1,\delta =0.5$). See also markers in phase diagram~\ref{fig:overview}(a) for the respective set of parameters.
	}
 \label{fig:results_time}
\end{figure}

As was discussed in Section \ref{sec:method}, the complex mode decomposition gives an advantage over direct evaluation of the diagrams when $\chi\sim\mathcal O(N^x)$ with $x<1$. We therefore investigating the dependence of the necessary number of modes $\chi_\epsilon(N)$  to acquire a given error threshold $\epsilon$  on $N$ time steps. The result is shown in  Fig.~\ref{fig:chi_vs_N}, using a threshold  $10^{-7}$ ($10^{-6}$) for finite (zero) temperature, see dotted lines in Fig.~\ref{fig:conv_test}.  For finite temperature, $\chi_\epsilon(N)$ saturates for large $N$, which is clearly a sublinear scaling with $N$. For zero temperature, such a saturation does not set in.  Nevertheless, $\chi$ is much smaller smaller than $N$ in absolute numbers, which gives orders of magnitude speedup for the third order integrals.  A power law  $\chi_\epsilon(N)\sim N^x$ cannot be extracted over the small range, but the data indicate a sublinear scaling in the asymptotic limit $N\rightarrow \infty$ even for the $T=0$ case. 

\subsection{Convergence of the strong-coupling series for $\mathcal{G}$}
\label{ssec:res_time}

From now on, we will focus on the low but finite temperature results ($T=1/50$), and analyze the convergence of the Green's function $\mathcal G^>_1(t)$.  We show two examples, in Fig.~\ref{fig:results_time}(a) close to the crossover between multi-boson and few-boson regime ($\Omega_0/\delta=2$, $\Delta E/\delta = 1.6$) and in Fig.~\ref{fig:results_time}(b) deep in the few-boson regime ($\Omega_0/\delta = 2$, $\Delta E/\delta = 0.5$). Deep in the few-boson regime, the numerical solutions converge to the exact one, order by order, such that the third order approximation already gives a reasonably accurate approximation for the full Green's function. Approaching the multi-boson regime, this is not the case and we observe a breakdown of the strong-coupling approximation. While the NCA and OCA solutions decay in time, the inclusion of third order diagrams in the self energy results in 
a Green's function which is increasing as $t\rightarrow \infty$. 
The exact result instead decays exponentially for large times, see black dotted line in Fig.~\ref{fig:results_time}(b). (The asymptotic behavior can be extracted from the analytic solution, see Eq.~\eqref{extdecay} in the Appendix.)
However, even though the strong-coupling series for $\mathcal{G}_1(t)$ at fixed order diverges at large $t$, one can observe a convergence of the $\mathcal{G}_1(t)$ with order at a fixed time. Hence, because of the finite core-hole lifetime $1/\Gamma$, one can in principle still obtain a convergent series for the XAS spectra \eqref{eq:ixas}, which will be analyzed in the following.

\subsection{X-ray absorption}
\label{ssec:res_xas}
%

\begin{figure*}
	\centering
	\begin{minipage}[c]{0.48\textwidth}
	
		\begin{minipage}{\textwidth}
		\raggedright (a)
		\centerline{\includegraphics[width=\textwidth]{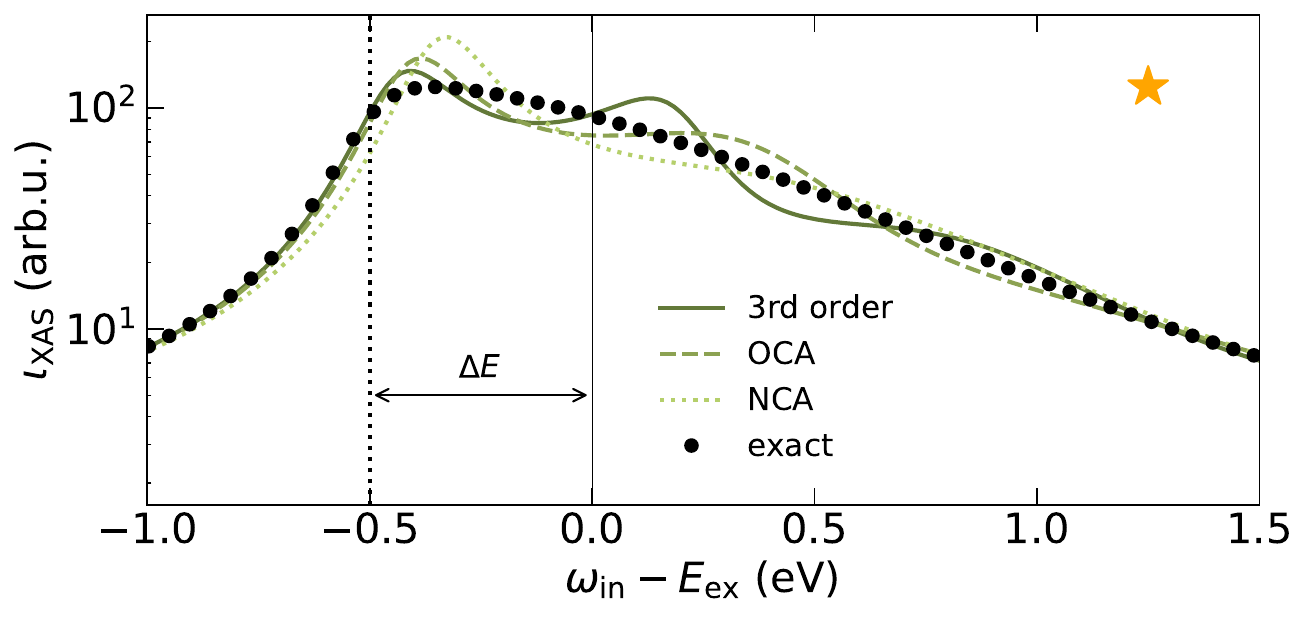}}
		\end{minipage} \\
		\begin{minipage}{\textwidth}
		\raggedright  (b)
		\centerline{\includegraphics[width=\textwidth]{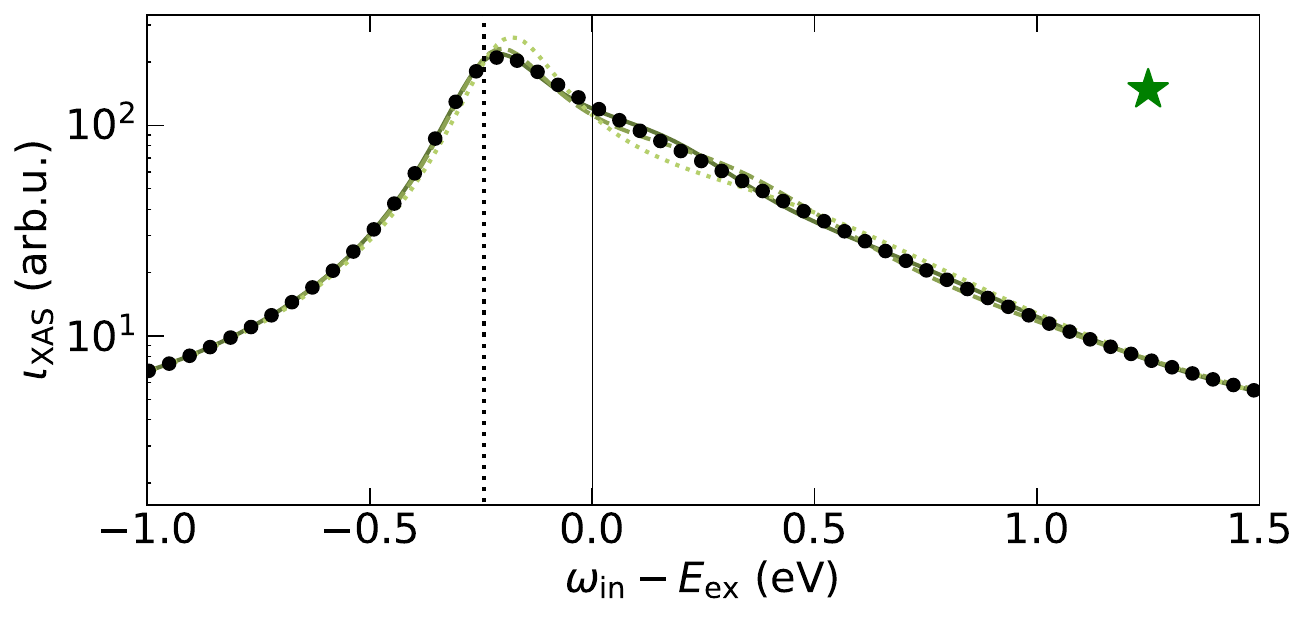}}
		\end{minipage}\\
				
	\end{minipage}%
	\hfill
	\begin{minipage}[c]{0.48\textwidth}
	
		\begin{minipage}{\textwidth}
		\raggedright  (c)
		\centerline{\includegraphics[width=\textwidth]{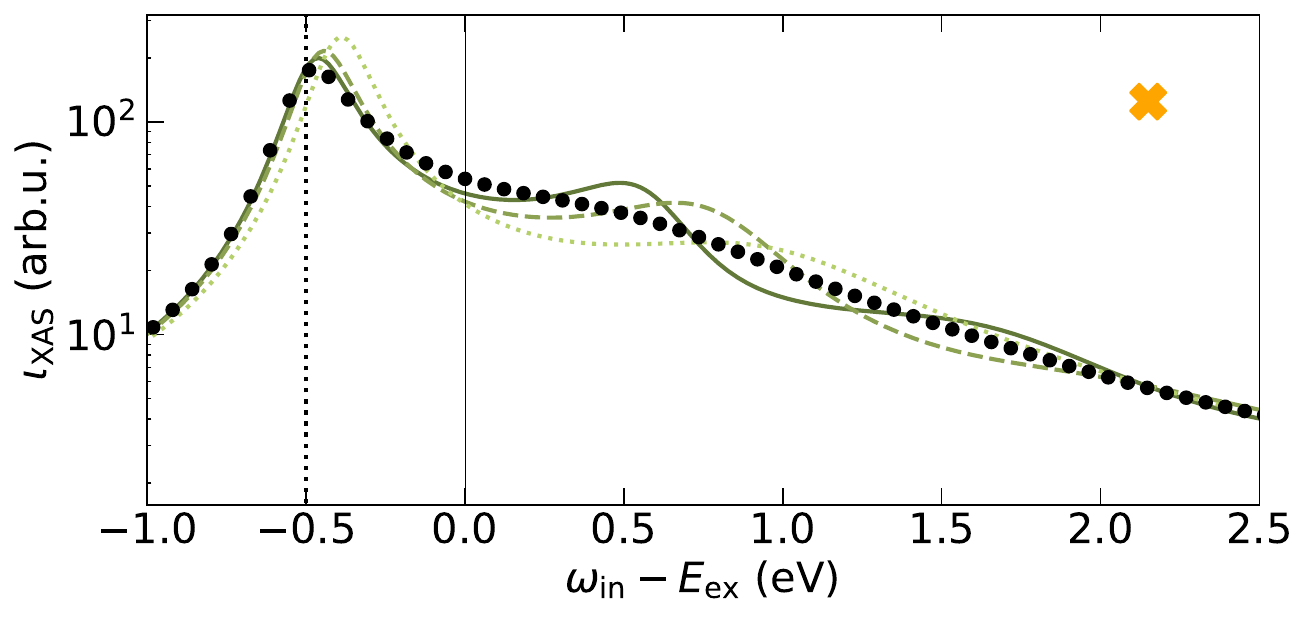}}
		\end{minipage} \\
		\begin{minipage}{\textwidth}
		\raggedright  (d)
		\centerline{\includegraphics[width=\textwidth]{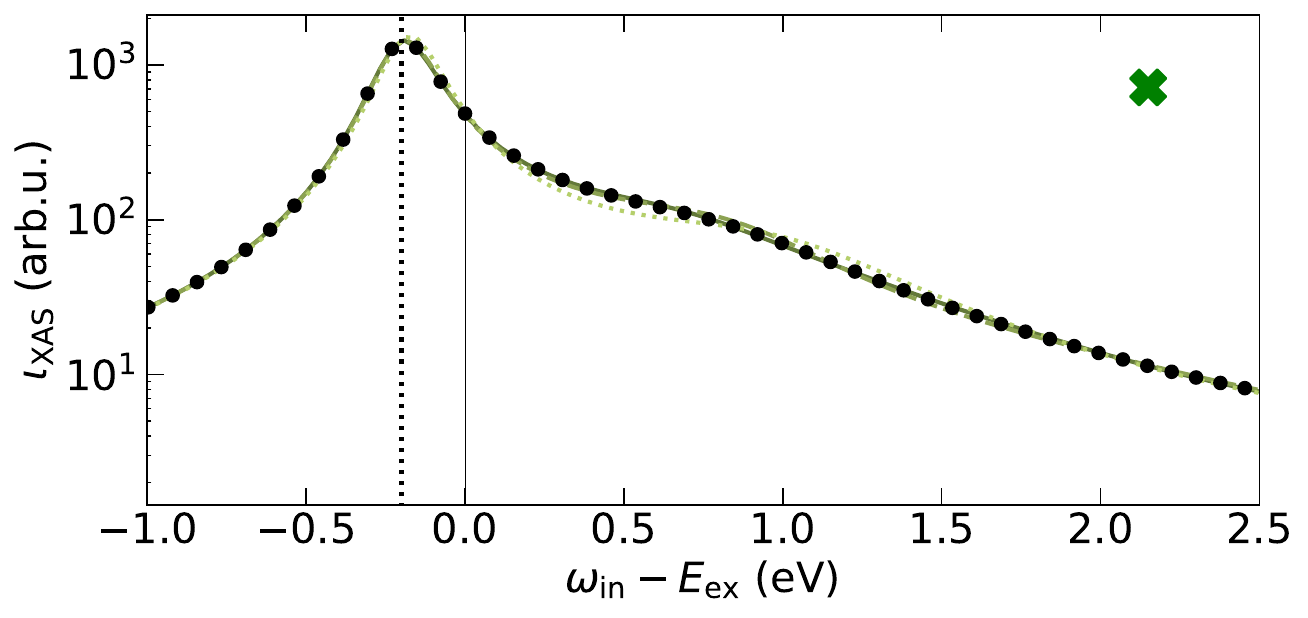}}
		\end{minipage}\\
		
	\end{minipage}
	
	\caption{XAS rate as a function of incoming photon energy according to the selected parameters in the phase diagram \ref{fig:overview}(a). The main absorption edge without screening is $E_{\text{ex}}$ (black solid line). The black dotted line gives the redshift \eqref{eq:redshift} for the exact result [black dots, Eq.~\eqref{eq:G_exact}].
The temperature is $T=1/50$, the inverse core hole lifetime is set to $\Gamma=0.1$ and $\chi=50$ modes have been used for the decomposition of the hybridization function.}
 \label{fig:results_XAS}
\end{figure*}

We will now examine the signatures of dynamic screening in the XAS spectra \eqref{eq:ixas}. (Throughout the analysis, the inverse core hole lifetime is set to $\Gamma=0.1$~eV.) First we discuss the exact results. Without screening, the XAS signal consists of a single peak at 
$\omega_{\rm in}= E_{\text{ex}}=\epsilon_d-\epsilon_c$ (vertical solid line) with a Lorentzian broadening $\Gamma$. The exact results with dynamic screening are shown by the black dotted lines in Fig.~\ref{fig:results_XAS}, where the left and right column represents the overdamped regime  ($\Omega_0/\delta < 1$) and the underdamped regime ($\Omega_0/\delta > 1$), respectively [see the star and cross markers in Fig.~\ref{fig:overview}(a)].
The dotted line at $\omega_{\rm in} - E_{\rm ex}=-\Delta E$ indicates the redshift $\Delta E$, which increases from bottom to top. In addition to the Lorentzian broadening, an asymmetric broadening arises due to excitations of screening modes. In the many-boson regime,  individual sidebands are not resolved, and multi-boson excitations generate a strong right shoulder of the absorption peak. Eventually,  the absorption maximum shifts from $\omega_{\rm in} =E_{\rm ex}-\Delta E$ to higher energies [see Fig.~\ref{fig:results_XAS}(a)].
The loss of distinct sidebands may also originate from a broad bosonic mode spectrum at large damping, even in the few-boson regime [Fig.~\ref{fig:results_XAS}(b)].

We now turn to analyze the convergence of the numerical results for the strong-coupling expansion.  At weak coupling (small $\Delta E$), already NCA and OCA results agree reasonable well with the analytic result, both in the weakly and strongly damped regime [Figs.~\ref{fig:results_XAS}(b,d)]. NCA underestimates the shift of the main absorption peak, which is systematically corrected by the OCA and third order. As the coupling is increased, the first signature of the breakdown of the strong-coupling expansion is the resolution of the satellite peaks [Figs.~\ref{fig:results_XAS}(a,c)]. The numerical results predict distinct sidebands, whereas the exact solution develops a shoulder-like structure where the individual multi-boson excitations are not resolved. 
Eventually, for large $\Delta E$ the strong-coupling expansion at low order does no longer give physical results. The red markers in the phase diagram \ref{fig:overview}(a) denote microscopic parameters where we could obtain a self-consistent solution for the Green's function in the NCA and OCA, but not in third order anymore. This phase diagram region is characterized by large coupling strengths $\gamma \gg \Omega_0,\delta$ such that the strong-coupling expansion in the interaction $\Delta\propto \gamma^2$ becomes ill-defined. 

\section{Conclusion and Discussion}
\label{sec:conclus}

In summary, this work has presented two main points:
(i) We have analyzed the quality of the strong-coupling expansion as an impurity solver for a model with retarded boson-mediated interaction.
(ii) In order to solve the strong-coupling series we have extended the impurity solver employing a decomposition of the time-nonlocal part in the action into decaying exponentials from imaginary time \cite{Kaye2024} to a real-time Keldysh contour. 

Regarding point (ii), we find that the method can lead to a substantial speedup of the numerical calculations. In particular, the computational complexity is mainly determined by the dimension of the decomposition and not by the number of time steps. Therefore, it represents a promising approach for intermediate and high order computations. We note that there is a relation to other decomposition schemes that use tensor cross interpolation (TCI) \cite{Fernandez2022,Fernandez2024,Erpenbeck2023b} for the evaluation of Keldysh diagrams \cite{Eckstein2024, Kim2024}. The exponential decomposition proposed in the present case is simpler in the sense that only a single time function must be decomposed instead of a multi-dimensional integrand-- even for more general cases in which the vertices would be operator-valued, e.g.~for a fermionic hybridization. Nonetheless, a rough estimate indicates that for the present problem the exponential decomposition would be similar in efficiency as compared to the TCI based approach (in particular since an additional gain in efficiency can be obtained when Prony's method is replaced by the more stable "matrix pencil method" \cite{Hua1990}). On the other hand, an important point to note is that the exponential decomposition is currently only implemented for the case where all vertices lie on a single contour branch ($\mathcal C_+$ or $\mathcal C_-$ in Fig.~\ref{fig:contour}). While this is sufficient for the present problem, in order to apply the technique to a generic impurity problem, it needs to be extended to allowing vertices on both contour branches. Though certainly possible, this is not straightforward and beyond the scope of this work.
The present analysis, however, already suggests that suitable extensions of the exponential decomposition approach can be a highly promising route for the solution of real-time impurity problems.

Regarding the point (i), our analysis shows that at not too large electron-boson coupling, low order corrections to NCA in the strong-coupling expansion provide a promising route to quantitatively discuss the effect of screening sufficiently. At larger couplings, instead, the expansion can become rather ill-behaved. It is therefore interesting to estimate the parameters relevant for realistic situations. In realistic $GW$+EDMFT calculations \cite{Golez2024}, the total electronic interaction is composed of the local Hubbard repulsion and the non-local screening. In the DMFT setting, latter is linked to the local interaction and the electronic hybridization through a self-consistency relation. When relating $GW$+EDMFT to the minimal model, the screened interaction $W$ can give a reasonable estimate for $\Delta$ \cite{Golez2024}. In Ref.~\cite{Golez2024}, $W$ was studied for a photo-doped charge transfer insulator. Compared to the interaction $W$ in equilibrium, photo-doping activates a broad spectral weight in the low-frequency region around $\Omega_0=1$eV. The coupling strength of these additional screening modes corresponds to redshifts $\Delta E$ up to $\approx 0.05 $eV for a few percent of photo-doping. Also, these screening modes are rather overdamped ($\delta \gtrsim\Omega_0$) due to their broad spectral weight. Given these effective parameters, one should place the photo-induced screening in the bottom left corner of the phase diagram \ref{fig:overview}(a), i.e.~in the overdamped, few-boson regime. This indicates that the strong-coupling expansion at low orders is a promising route for a quantitative analysis of XAS spectra in realistic photo-doped systems. For a full solution of the DMFT impurity problem, both fermion and boson environments need to be considered, which could be achieved at high orders using a suitable extension of the exponential decomposition method (see discussion above).

\acknowledgments
We thank Denis Gole\v{z}, Benoît Richard and Philipp Werner for helpful discussion on the topic. Both authors are supported by the Cluster of Excellence „CUI: Advanced Imaging of Matter“ of the Deutsche Forschungsgemeinschaft (DFG) – EXC 2056 – project ID 390715994.

\appendix

\section{Analytic expression for the Green's function $\mathcal G^>_1$}
\label{app:ppgf_exact}

In the following we give the derivation of Eq.~\eqref{eq:G_exact} in the main text. The key step is to apply a multi-mode Lang-Firsov transform \cite{LangFirsov1962} to the Hamiltonian\eqref{eq:LF_Hamiltonian-1}. We will drop spin-dependency and assume a spinless fermion, since the valence orbital has a fermionic occupation $n_d$ of at most one (cf.~Section \ref{ssec:XAS_model_S}). Creation and annihilation operators of two different modes commute with each other. Therefore, we can perform a Lang-Firsov transform on all couplings simultaneously,
\begin{alignat}{2}
	\tilde H_d &= e^{\mathcal{S}} H_d e^{-\mathcal{S}} &&= \epsilon_d n_d + \sum_\alpha \omega_\alpha b^\dagger_\alpha b_\alpha - \gamma^2 \sum_\alpha \frac{n_d^2}{\omega_\alpha} \nonumber \\
	& \phantom{0} &&= \tilde \epsilon n_d + \sum_\alpha \omega_\alpha b^\dagger_\alpha b_\alpha ,
	\label{eq:H_tilde} \\
	\tilde d &= e^{\mathcal{S}} d e^{-\mathcal{S}} && = d \exp\left[-\sum_\alpha \frac{\gamma}{\omega_\alpha} (b_\alpha-b^\dagger_\alpha)\right] ,
	\label{eq:c_tilde}
\end{alignat}
where $\mathcal{S}= \sum_\alpha (n_d \gamma/\omega_\alpha) (b_\alpha -b^\dagger_\alpha)$ is the generator of the transform. For spinless fermions, $n_d^2 = n_d$, such that the final Hamiltonian $\tilde H_d$ describes a set of uncoupled oscillators with frequencies $\{\omega_\alpha\}_\alpha$ and a fermionic level at renormalized energy $\tilde \epsilon = \epsilon_d-\gamma^2 \sum_\alpha (1/\omega_\alpha)$. In the continuum limit this shift becomes $-\gamma^2 \sum_\alpha (1/\omega_\alpha) \rightarrow -\gamma^2 \int_0^\infty \mathrm d\omega \mathcal A(\omega) / \omega$, with $\mathcal A$ the density of states of the oscillator continuum. To relate $\mathcal A$ to the spectral function  $A_{\text{b}}$ [Eq.~\eqref{eq:A_Delta}], we note that the Green's functions $C_{X,\alpha}(t) = -i \langle \mathrm T_{\mathcal C} X_\alpha(t) X_\alpha \rangle $ and $G_\alpha(t) = -i \langle \mathrm T_{\mathcal C} b_\alpha(t) b_\alpha^\dagger \rangle $ are related via $C^R_{X,\alpha}(t) = \text{Re}\,G_\alpha^R(t) $, with $G_\alpha^R(t)=-i\theta(t) e^{-i\omega_\alpha t}$. This implies
\begin{align}
	A_{\text{b}}(\omega) 
	&=  -\frac{1}{\pi} \sum_\alpha \text{Im } \int_{-\infty}^\infty \mathrm dt \, C^R_{X,\alpha}(t)  e^{i\omega t} \nonumber  \\
	&= \frac{1}{2} \sum_\alpha \left[ \delta(\omega-\omega_\alpha)-\delta(\omega+\omega_\alpha) \right] \nonumber \\
	&=
	\frac{1}{2} \int_0^\infty \mathrm d\omega_\alpha \, \mathcal A(\omega_\alpha)  \left[ \delta(\omega-\omega_\alpha)-\delta(\omega+\omega_\alpha) \right] 
	\nonumber \\
	&= \frac{\text{sgn}(\omega)}{2} \mathcal A(|\omega|).
\end{align}
In particular, $\mathcal A(\omega) = 2 A_{\text{b}}(\omega)$ for $\omega>0$.

Now, according to definition \eqref{ppG1}, the greater pseudo-particle Green's function of the final X-ray excited state is 
\begin{align}
	\mathcal G^>_1 (t'-t)
	&= -i \langle \sigma_d\bar\sigma_c | \mathrm T_{\mathcal C} e^{iS(t'\succ t)} | \sigma_d\bar\sigma_c \rangle \nonumber \\
	&=	-i \,\text{tr} \left[ \rho \,e^{i H t'} \,c^\dagger_\sigma  d_\sigma \, e^{-i H (t'-t)} d^\dagger_\sigma c_\sigma \, e^{-i H t} \right] 	, \nonumber \\
	&= 		-i \,\text{tr} \left[ \rho \, c^\dagger_\sigma (t') d_\sigma (t') \, d^\dagger_\sigma(t)  c_\sigma (t) \right]	,
	\label{eq:aux}
\end{align}
where $t'\succ t$, such that time propagation runs from $0_+\to t\rightarrow t'\to 0_-$ along the Keldysh contour (see Fig.~\ref{fig:contour}).
The Hamiltonian $H=H_d+H_c$ is composed of the valence and core part (cf.~Sec.~\ref{ssec:XAS_model_H}).
As given in the main text, $\ket{\sigma_d\bar\sigma_c} = d^\dagger_\sigma c_\sigma \ket{0_d 2_c}$ is the final X-ray excited state with one electron in the valence orbital, and $\ket{0_d 2_c}$ denotes the initial state before the X-ray probe. In the Hamiltonian formalism where we include the bosonic degrees of freedom into the Hilbert space explicitly, this corresponds to the initial joint density matrix $\rho= \ket{0_d 2_c}\bra{0_d 2_c} \otimes \rho_{\text{therm}}$, with $\rho_{\text{therm}}= \exp(-\beta\sum_\alpha \omega_\alpha b^\dagger_\alpha b_\alpha)/Z$ the thermal density matrix of a bosonic ensemble.
Evaluating expression \eqref{eq:aux} in the Lang-Firsov-transformed basis yields 
\begin{align}
	&\mathcal G^>_1 (t)  = -i \,\text{tr} \big[ \tilde \rho \, e^{-\sum_\alpha \frac{\gamma}{\omega_\alpha}(b_\alpha(t)-b_\alpha^\dagger(t)) } c^\dagger_\sigma (t) d_\sigma(t) \, 
	\nonumber
	\\
	&\,\,\,\,\,\,\,\,\,\,\,\,\,\,\,\,\,\,\,\,\,\,\,\,\,\,\,\,\,\,\,\,\,\,\,
	e^{\sum_\alpha \frac{\gamma}{\omega_\alpha}(b_\alpha-b_\alpha^\dagger) } d_\sigma^\dagger c_\sigma\big] 
	\nonumber \\
	&=	-i e^{-i(\tilde\epsilon + \epsilon_c) t} \, \text{tr} \left[  \rho_{\text{th}} \, e^{-\sum_\alpha \frac{\gamma}{\omega_\alpha}(b_\alpha(t)-b_\alpha^\dagger(t)) } e^{\sum_\alpha \frac{\gamma}{\omega_\alpha}(b_\alpha-b_\alpha^\dagger) } \right],
	\label{eq:therm_avg}
\end{align}
using that $\ket{0_d 2_c}$ and $\rho_{\text{th}}$ are not affected by the transform because $n_d=0$ in the initial state, hence $\rho=\tilde\rho$. We will now focus on the evaluation of the remaining average which can be achieved by Wick's theorem. 
Define the operators 
\begin{align}
	\begin{split}
	A
		&= -\sum_\alpha \frac{\gamma}{\omega_\alpha}(b_\alpha(t)-b_\alpha^\dagger(t)) \\
		&= -\sum_\alpha \frac{\gamma}{\omega_\alpha}(e^{-i\omega_\alpha t} b_\alpha - e^{i\omega_\alpha t} b_\alpha^\dagger),
	\end{split}
	\\
	B
		&= \sum_\alpha \frac{\gamma}{\omega_\alpha}(b_\alpha-b_\alpha^\dagger),
\end{align}
which are linear combinations of annihilation and creation operators, such that $[A,[A,B]]=[B,[A,B]]=0$ because $[A,B]$ is proportional to the identity operator. Employing the Baker-Campbell-Hausdorff formula, we find $e^A e^B = e^{A+B+\frac{1}{2}[A,B]} = e^{A+B} e^{\frac{1}{2}[A,B]} $. To apply Wick's theorem to the operator $e^{A+B}$, we first expand the exponential and find that only the even terms contribute because the expectation value has to contain the same number of annihilation and creation operators,
\begin{align}
	\text{tr} \left[ \rho_{\text{th}} e^{A+B} \right]
	&=   \sum_{n=0}^\infty \frac{\text{tr} \left[ \rho_{\text{th}} (A+B)^{2n} \right] }{(2n)!} \\
	&= \sum_{n=0}^\infty \frac{(2n-1)!! \, \text{tr} \left[ \rho_{\text{th}} (A+B)^{2} \right]^n }{(2n)!} 
\end{align}
In the second line Wick's theorem is inserted, by noting that there exist $(2n-1)!! = \prod_{j=1}^n (2j-1)$ pairings for $2n$ operators, where the order within the pair does not matter. Further simplifying $(2n)! / (2n-1)!!   = 2^n n!$ and collecting the terms, one finds
\begin{align} \begin{split}
	&\text{tr} \left[ \rho_{\text{therm}}e^A e^B \right] \\
	&\quad = \exp \text{tr} \left\{ \rho_{\text{th}} \frac{1}{2} \left(A^2 + B^2 + AB +BA + [A,B]\right) \right\} \\
	&\quad = \exp \text{tr} \left[ \rho_{\text{th}} \frac{1}{2} \left(A^2 + B^2 + 2AB\right) \right].
\end{split}\end{align}
A straightforward calculation yields
\begin{align}
	\text{tr} \left[ \rho_{\text{th}} A^2 \right] &= -\sum_\alpha \frac{\gamma^2}{\omega_\alpha^2} \text{tr} \left[ \rho_{\text{therm}} \left(b_\alpha b_\alpha^\dagger + b^\dagger _\alpha b_\alpha\right) \right] \\
	&= \text{tr} \left[ \rho_{\text{th}} B^2 \right],
\end{align}
\begin{align}
	\begin{split}
	&\text{tr} \left[ \rho_{\text{th}} AB \right] \\
	& \quad = -\sum_\alpha \frac{\gamma^2}{\omega_\alpha^2} \text{tr} \left[ \rho_{\text{th}} \left(e^{-i\omega_\alpha t} b_\alpha b_\alpha^\dagger +e^{i\omega_\alpha t}  b^\dagger _\alpha b_\alpha\right) \right],
	 \end{split}
\end{align}
such that
\begin{align} \begin{split}
	&\text{tr} \left[ \rho_{\text{th}} \frac{1}{2} \left(A^2 + B^2 + 2AB\right) \right]  \\
	&\quad=
	\sum_\alpha \frac{\gamma^2}{\omega_\alpha^2} \left[ (e^{-i\omega_\alpha t}-1) \overline b(\omega_\alpha) + (e^{i\omega_\alpha t}-1) b(\omega_\alpha)\right] \\
	&\quad= : f^>(t),
\end{split}
\end{align}
with $b(\omega)= 1/(e^{\beta \omega}-1)$ the Bose distribution function and $\overline b(\omega)= 1+b(\omega)=-b(-\omega)$.
In the thermodynamic limit this expression becomes
\begin{align}
	f^>(t) &= \gamma^2 \int_0^\infty\mathrm d\omega \, \frac{\mathcal A(\omega) }{\omega^2} \left[ (e^{-i\omega t}-1) \overline b(\omega) + (e^{i\omega t}-1) b(\omega)\right] \\
	&= \gamma^2 \int_{-\infty}^\infty \mathrm d\omega \,  \frac{2 A_{\text{b}}(\omega)}{\omega^2} (e^{i\omega t} -1) b(\omega),
	\label{eq:f>}
\end{align}
employing the relation between $b(\omega)$ and $\overline b(\omega)$ and the odd parity of the spectral function, $A_{\text{b}}(-\omega)=-A_{\text{b}}(\omega)$, in the second line.
With that we summarize
\begin{align}
	\mathcal G_1^>(t) = -i e^{-i(\tilde \epsilon+\epsilon_c) t} e^{f^>(t)},
	\label{eq:G_exact_app}
\end{align}
with energy level $\tilde \epsilon = \epsilon_d -\gamma^2 \int_0^\infty \mathrm d\omega \,2 A_{\text{b}}(\omega) / \omega$ and function $f^>(t)$ given in \eqref{eq:f>}.
Note that for the hybridization \eqref{eq:A_Delta} representing a damped oscillator, the integrand in \eqref{eq:f>} diverges like $\sim 1/\omega$ for $|\omega|\ll 1$ such that the integral has to be treated as principal value and evaluated accordingly.

For the bosonic spectrum \eqref{Abos} given in the main text and employed throughout this paper, we can calculate the redshift $\Delta E = \epsilon_d - \tilde\epsilon$ via partial fraction decomposition and complex contour integration,
\begin{align}
	\Delta E = \frac{\gamma^2 \Omega_0}{\Omega_0^2 + \delta^2}.
	\label{eq:redshift_model}
\end{align}

Lastly, we are interested in the behavior of $\mathcal{G}^>_1$ at asymptotically large times $t$ for finite temperature ($T>0$). We show that $\mathcal G^>_1(t)$ decays exponentially at large times like 
\begin{align}
\mathcal G^>_1(t)\sim \exp\left(-4\frac{\gamma^2\Omega_0\delta}{\beta(\delta^2+\Omega_0^2)^2} t\right). 
\label{extdecay}
\end{align}
This can be seen from the analytic expression \eqref{eq:G_exact_app} of the Green's function $\mathcal G^>_1(t)$: $\tilde\epsilon,\epsilon_c \in\mathbb R$ and therefore any decay in time has to originate from the the real part of $f^>(t\rightarrow\infty)$. For $t\rightarrow \infty$ we split the integral defining $f^>(t)$ as follows,
\begin{align}
	f^>(t) = \gamma^2 \left[ \int_{-\infty}^\infty \mathrm d\omega\, \frac{2 A_{\text{b}}}{\omega^2} b(\omega) e^{i\omega t} - \int_{-\infty}^\infty \mathrm d\omega \, \frac{2 A_{\text{b}}}{\omega^2} b(\omega) \right],
\end{align}
where each integral by itself is ill-defined. However, their sum remains finite because the second term cancels divergencies of the first at $t\rightarrow 0$. For long times, we can therefore ignore the second part and focus on the Fourier integral. Its long-time behavior is determined by the dominant scaling of the integrand close to the singular region, i.e.~$\sim1/\omega^2$ for small $\omega$ and the Fourier transform of $1/\omega^2$ gives $-\pi|t|$. 
In total, this yields the exponential decay \eqref{extdecay} of $\mathcal G^>_1(t)$.

\section{Pronys method}
\label{ssec:cplx_mode_decomp}

In this section we summarize Pronys method for the construction of a complex mode decomposition. For a good introduction into the method we refer to Refs.~\onlinecite{FernandezRodriguez2018,Sauer2018} and \onlinecite{Makhoul1975} for obtaining the linear difference equation.

A function $F:\mathbb R\to \mathbb C$, sampled at the equidistant points $n\delta t$, with $n=0,\dots,N$, defines a discrete signal $F_n = F(n\,\delta t)$. We want to find an approximation $\tilde F_n$ for $F_n$, such that
\begin{align}
	\tilde F_n = \sum_{m=1}^\chi b_m e^{\omega_m n\delta t} = \sum_{m=1}^\chi b_m z_m^{n}.
	\label{eq:exp_decomp_aim}
\end{align}
The number of \textit{complex}-valued modes $\omega_m$ (or, \textit{complex} roots $z_m$) is $\chi$, which can be chosen arbitrary except that $\chi<N$. The advantage of this decomposition over a Fourier series is that (i) this decomposition is well-behaved when applied to decaying signals, (ii) the modes $\omega_m$ are not determined by the time grid, but optimally chosen to describe the signal $F_n$.

The key idea of the algorithm is linked to the observation that \textit{if} $\tilde F_n$ is the solution of a linear difference equation,
\begin{align}
	\tilde F_n = -\sum_{p=1}^\chi c_p \tilde F_{n-p}, \quad \text{for } n=\chi,\chi+1,\dots,N,
	\label{eq:lin_diff_eq}
\end{align}
then it can be written as a linear combination of exponential functions with complex frequencies, as desired \eqref{eq:exp_decomp_aim}. This is due to the fact that $\tilde F_n =e^{n\omega\delta t}\equiv z^n$ will be a solution of \eqref{eq:lin_diff_eq},
if and only if $z= \exp(\omega\delta t)$ is one of the $\chi$ roots of the corresponding characteristic polynomial to the difference Eq.~\eqref{eq:lin_diff_eq}, i.e., 
\begin{align}
P(z)=z^\chi+\sum_{p=1}^\chi c_p z^{\chi-p}=0.
\end{align}
Then the decomposition \eqref{eq:exp_decomp_aim} is simply a decomposition of the signal $\tilde F_n$ into a non-orthonormal basis of solutions for the linear difference eq.~\eqref{eq:lin_diff_eq}.

Hence, the first step is to find the difference equation for the original signal $F_n$. This is closely related to a scheme called Linear Prediction, which is well known in signal processing \cite{Makhoul1975}. Its name stems from the fact that a difference equation can be used to predict the signal at future times which are not measured yet. We will not discuss the existing literature in the field of Linear Prediction, but instead give a summary of the steps involved to obtain a complex mode decomposition:

(1) \textit{Linear difference equation}: Find a inhomogeneous linear difference equation for the original signal $F_n$,
\begin{align}
	F_n = -\sum_{p=1}^\chi c_p F_{n-p} + r_p, \quad \text{for } n=\chi,\chi+1,\dots,N,
	\label{eq:lin_diff_eq_approx}
\end{align}
by minimizing the residuals errors $r_p$. This can be done in various ways. For example, one can minimize the total squared error $\sum_{n=\chi}^N \text{err}_n^2$, with $\text{err}_n=F_n -(-\sum_{p=1}^\chi F_{n-p})$, which yields the normal equation $-\sum_{p'=1}^\chi c_{p'}\Phi_{p'p}=\Phi_{0p}$ for $p=1,2,\dots,N$, with $\Phi_{p'p}=\sum_{n=\chi}^N F_{n-p'} F_{n-p}$ being the autocorrelation matrix. Alternatively, if $N \geq 2\chi-1$,  Eq.~\eqref{eq:lin_diff_eq_approx} can be rewritten as an overdetermined system of linear equations,
\begin{align}
		    	- \begin{pmatrix} F_{\chi} \\ F_{\chi+1}\\ \vdots\\ F_N \end{pmatrix}
		    	   =
		    	\begin{pmatrix}
		    		F_{\chi-1} & F_{\chi-2} & \dots & F_{0}\\ 
		    		F_{\chi} & F_{\chi-1}	& \dots & F_1 \\
		    		\vdots & \vdots & & \vdots \\
		    		F_{N-1} & F_{N-2} & \dots & F_{N-\chi}\\
		    	\end{pmatrix}
		    	\begin{pmatrix}
		    		c_1 \\ c_2 \\\vdots \\ c_\chi
		    	\end{pmatrix},
		    	\label{eq:linearization}
\end{align}
whose coefficients $c_p$ can be determined based on a least square fit. This is formally equivalent to the normal equation approach and can be tackled by singular-value-decomposition.

(2) \textit{Characteristic polynomial}: Having found the coefficients $c_p$ of the linear difference equation, one solves for the roots of the corresponding characteristic polynomial $z^\chi+\sum_{p=1}^\chi c_p z^{\chi-p} =: a_0 +a_1 z+a_2z^2+\dots + a_{\chi} z^{\chi}$. One straightforward approach to obtain these is by calculating the eigenvalues of the companion matrix \cite{NumRecipesC}
\begin{align}
			   	\begin{pmatrix}
		    		-\frac{a_{\chi-1}}{a_\chi} & -\frac{a_{\chi-2}}{a_\chi} & \dots  & -\frac{a_{2}}{a_\chi} & -\frac{a_{1}}{a_\chi} & -\frac{a_{0}}{a_\chi}\\ 
		    		1 & 0	& \dots & 0 & 0 & 0 \\
		    		0 & 1	& \dots & 0 & 0 & 0 \\
		    		\vdots & \vdots & & \vdots & \vdots & \vdots \\
		    		0 & 0	& \dots & 1 & 0 & 0 \\
		    		0 & 0	& \dots & 0 & 1 & 0 \\
		    	\end{pmatrix}.
			\label{eq:companion_matrix}
\end{align}
The eigenvalues of this matrix will be the roots of the characteristic polynomial. This can be seen from Vieta's formulas.

(3) \textit{Coefficient fit}: Finally, compute the coefficients $b_m$ in the complex mode decomposition \eqref{eq:exp_decomp_aim}.	Here, again, we can reduce the problem to a least square fit of an overdetermined system of linear equations. This is achieved by defining the matrix $Z\in \mathbb C^{(N+1)\times \chi}$ with matrix elements $Z_{nm}=z^n_m$. Then, in terms of the original signal $F_n$, the complex mode decomposition \eqref{eq:exp_decomp_aim} becomes $F_n = \sum_{m=1}^\chi Z_{nm} b_m$. This is solved for the best fit of coefficients $b_m$.


%

\end{document}